\documentclass[twocolumn,reprint,superscriptaddress, amsmath,amssymb, floatfix, aps, prb]{revtex4-2}

\usepackage{graphicx,subfigure}
\usepackage{enumerate}
\usepackage{color}
\usepackage{dcolumn}
\usepackage{epstopdf}
\usepackage{multirow}
\usepackage{appendix}
\usepackage{braket} 
\usepackage{cancel}
\usepackage{amsfonts}
\usepackage{amsmath}
\usepackage{amssymb}
\usepackage{mathrsfs}
\usepackage{epsfig}
\usepackage{pict2e}
\usepackage{bm}
\usepackage[normalem]{ulem}

\usepackage{colortbl} 

\usepackage[colorlinks, breaklinks, citecolor=blue, linkcolor=blue,  urlcolor={blue}]{hyperref}

\definecolor{mygray}{gray}{.9}
\definecolor{darkblue}{rgb}{1,1,.70}
\definecolor{lightblue}{rgb}{1,1,.90}

\makeatother

\begin{document}

\title{Geometric Bloch oscillations and transverse displacement in flat band systems}

\author{Jing-Xin Liu}
\email[Email: ]{602023220045@smail.nju.edu.cn}
\affiliation{National Laboratory of Solid State Microstructures and School of Physics, Nanjing University, Nanjing 210093, China}
\affiliation{Collaborative Innovation Center of Advanced Microstructures, Nanjing 210093, China}

\author{Giandomenico Palumbo}
\email[Email: ]{giandomenico.palumbo@gmail.com}
\affiliation{School of Theoretical Physics, Dublin Institute for Advanced Studies, 10 Burlington Road, Dublin 4, Ireland}

\author{Marco Di Liberto}
\email[Email: ]{marco.diliberto@unipd.it}
\affiliation{Dipartimento di Fisica e Astronomia “G. Galilei” and Padua Quantum Technologies Research Center,
Universit\'a degli Studi di Padova, 35131, Padova, Italy}
\affiliation{Istituto Nazionale di Fisica Nucleare (INFN), Sezione di Padova, 35131 Padova, Italy}

\begin{abstract}
We investigate transport phenomena and dynamical effects in flat bands where the band dispersion plays no role. 
We show that wavepackets in geometrically non-trivial flat bands can display dynamics when inhomogeneous electric fields are present. 
This dynamics is revealed both for the wavepacket trajectory and for its variance, for which we derive semiclassical equations extended to the non-Abelian case. 
Our findings are tested in flat band models in one- and two-dimensional lattices where the dynamics is solely determined by geometric effects, in the absence of band dispersion.
In particular, in the one-dimensional case, we show the existence of Bloch oscillations for the wavepacket position and for the wavepacket variance, whereas in the two-dimensional case we observe a transverse displacement of the wavepacket in the absence of Berry curvature.
This work paves the way for understanding quantum-geometry-induced dynamical effects in flat band materials and also opens the possibility for their observation with synthetic matter platforms. 
\end{abstract}

\maketitle

\section{Introduction}
\label{sec:1}

The collective behavior of electrons in a crystal is described by Bloch electrons and their transport properties rely on the intrinsic topological and dispersive properties of the Bloch bands \cite{Girvin}.
In this respect, phenomena like the anomalous ~\cite{Nagaosa-RevModPhys.82.1539} and integer quantum Hall effect ~\cite{Haldane-PhysRevLett.61.2015}, 
originate from Berry phase effects of the bands \cite{Niu-RevModPhys.82.1959}. 
The band structure of lattice systems also give rise to other phenomena, such as Bloch oscillations \cite{Girvin}, which are intrinsically related to the band dispersion. 
These manifest via a periodic oscillation of an electron wavepacket in real and momentum space when a homogeneous electric field is applied in the absence of dissipation.
However, a non-trivial interplay between topology, band dispersion and crystalline symmetries has been shown to give rise to topological Bloch oscillations~\cite{Alexandradinata}. 
These are Bloch oscillations displaying a periodicity that is an integer multiple of the fundamental period~\cite{Alexandradinata}, whose protection relies on the topological properties of the bands. 
They have been observed with cold atoms \cite{Li2016} and identified in higher-order topological insulators~\cite{Marco-ncs41467-020-19518-x}.

More recently, band geometry has emerged as a novel and central paradigm in condensed matter physics. 
It is based on the concept of quantum metric \cite{Provost}, namely a momentum-space Riemannian metric, which has already found a large number of applications in quantum matter~\cite{Marzari-PhysRevB.56.12847,Matsuura-PhysRevB.82.245113,Neupert-PhysRevB.87.245103,Kolodrubetz-PhysRevB.88.064304,Legner-PhysRevB.88.115114,MaYuquan-PhysRevE.90.042133,Gao-PhysRevLett.112.166601,Gao-PhysRevB.91.214405,Srivastava-PhysRevLett.115.166802,Piechon-PhysRevB.94.134423,Freimuth-PhysRevB.95.184428,Lianglong-PhysRevB.96.064511,Ozawa-PhysRevB.97.041108,Palumbo2017,Palumbo2018,Ozawa-PhysRevB.97.201117,Bleu-PhysRevLett.121.020401,Salerno2020,Zhu2021,Mera-PhysRevB.106.165133,Lin2024,Jankowski2025,Oancea2025}.
In this context, linear external fields have shown to provide nonlinear responses~\cite{Jiang2025}, such as the nonlinear Hall effect~\cite{Fu-PhysRevLett.115.216806,Gao-PhysRevLett.112.166601,Gao-doi:10.1126/science.adf1506,Kamal-PhysRevB.108.L201405,kang2019nonlinear,kumar2021room,lai2021third,du2021nonlinear,li2024quantum,Jain2024} and circular photogalvanic effect~\cite{Juan2017}, based on the Berry curvature. 
However, other nonlinear optical responses and quantum effects~\cite{Ahn-2022NatPhys,Jankowski2024,Jankowski2025-2} have been shown to be related to band geometry \cite{PhysRevLett.132.106601,PhysRevLett.132.196801,PhysRevB.105.085154,PhysRevLett.134.116301,PhysRevLett.133.226302,PhysRevB.110.195437,PhysRevB.109.115121,PhysRevB.108.L140503,PhysRevLett.130.266601,PhysRevB.111.165130,PhysRevB.111.L121102}.
These phenomena can be understood through the semiclassical equations of motion (EOM) of wavepackets~\cite{Niu-RevModPhys.82.1959}, which offer an intuitive perspective on both topological and geometrical properties of bands~\cite{Niu-PhysRevLett.75.1348,Niu-PhysRevB.53.7010,Moore-PhysRevB.87.165110,Moore-PhysRevLett.105.026805,Moore-PhysRevLett.115.117403}. 
\begin{figure}[!b]
\centering
\includegraphics[width=8.5cm]{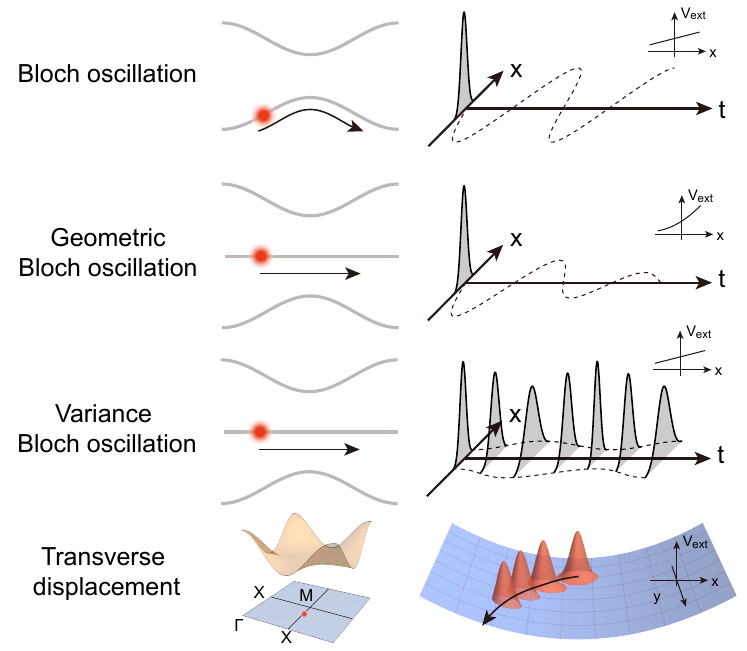}
\caption{Dynamical phenomena for wavepackets. In the first panel, conventional Bloch oscillations are shown to arise from dispersive bands in the presence of linear electric fields. 
In the second and third panels, we represent geometric Bloch oscillations and variance Bloch oscillations originating from the intrinsic quantum geometry of flat bands with external inhomogeneous potential. 
In the last panel, we represent a geometric induced transverse displacement.}
\label{fig:GBO}
\end{figure}
Based on this semiclassical description, strategies employing wavepacket dynamics have been developed for ultracold atoms in optical lattices, for instance, to extract the Chern number~\cite{Price-PhysRevA.85.033620,Goldman_2014,Goldman-PhysRevLett.111.135302,Goldman-2016Nat.Phys,Mugel-2017SciPostPhys,Zhu-2018Adv.in.Phys, Aidelsburger-2015Nat.Phys,Jotzu-2014Nature,Wintersperger-2020Nat.Phys}, the Zak phase \cite{Atala-2013Nature} and the Wilson loops \cite{Duca-Science} of topological Bloch bands. 
Furthermore, these approaches have also been successfully employed to analyze synthetic classical systems \cite{Raghu-PhysRevA.78.033834}, including photonic ones \cite{Ozawa-RevModPhys.91.015006}.

The analysis of the EOM for wavepackets in \emph{inhomogeneous} external electric and magnetic fields has shown an enhanced role of band geometry for quantum transport phenomena~\cite{Lapa-PhysRevB.99.121111,Kozii-PhysRevLett.126.156602, Northe2022, Ali-PhysRevResearch.7.013141,Xu2025}.
For instance, some relevant quantized observables in topological matter, such as the Hall viscosity, can be detected through inhomogeneous electric fields~\cite{Hoyos-PhysRevLett.108.066805,Bradlyn-PhysRevB.86.245309,Harper}. 
As discussed in detail in Refs.~\cite{Lapa-PhysRevB.99.121111,Kozii-PhysRevLett.126.156602}, an inhomogeneous electric field provides a contribution to the wavepacket velocity including the quantum metric and a rank-3 tensor \cite{Kozii-PhysRevLett.126.156602} up to cubic-order terms in the electric field.  
Importantly, band geometry does not only influence the dynamics of the wavepacket mean position, but also the adiabatic evolution of its shape.
In fact, because the variance of a maximally localized Wannier function is related to the trace of the quantum metric~\cite{Marzari-PhysRevB.56.12847}, it has been shown~\cite{Souza-PhysRevB.79.045127} that the variance of the wavepacket has non-trivial dynamics due to the quantum metric already in the presence of homogeneous electric fields.

Systems with flat and nearly-flat bands are a promising platform for exploring the role of quantum geometry, as for ideal fractional Chern insulators~\cite{Roy2014,Sondhi2013,Thomale2017,Mera2021}. 
Experimental progress in realizing ideal fractional Chern insulators in twisted graphene and $\mathrm{MoTe_2}$ systems~\cite{park2023observation,cai2023signatures,xie2021fractional} has opened new avenues for investigating quantum geometry in strongly correlated electron systems. 
Besides, the superfluid weight (or stiffness) in flat bands has been theoretically shown to originate from quantum metric effects~\cite{peotta2015superfluidity,PhysRevB.104.144507,PhysRevLett.127.170404,PhysRevLett.131.240001,PhysRevLett.128.087002,PhysRevB.109.214518,PhysRevB.107.224505,PhysRevLett.132.026002,PhysRevA.107.023313,PhysRevB.106.184507,PhysRevA.105.023301,PhysRevB.111.134511}. 
Thus, flat and nearly flat bands represent an ideal playground in which it is possible to investigate new physical phenomena induced by the band geometry.

In this work, we unveil dynamical effects that arise from the non-trivial interplay 
of band geometry and inhomogeneous electric fields in flat bands.
We derive the semiclassical EOM for wavepackets in both degenerate and non-degenerate band models in the presence of inhomogeneous electric fields up to the cubic order. 
We apply those equations to several flat-band models in both one and two dimensions.
Contrary to the common intuition concerning semiclassical dynamics in flat bands, we crucially show the emergence of a \emph{geometrically-induced Bloch oscillation} of the wavepacket position and variance, as represented in Fig.~\ref{fig:GBO}. 
In particular, we stress the importance of quantum metric effects in understanding the wavepacket variance evolution, thus generalizing the results obtained in Ref.~\cite{Souza-PhysRevB.79.045127}.
Furthermore, in two dimensions we show a transverse displacement of the wavepacket without Berry curvature.

This paper is organized as follows.
In Sec.~\ref{sec:2}, we present the equations for the mean position and variance of the wavepacket in both the real and momentum space.
In Sec.~\ref{sec:3}, we numerically validate our results in flat-band Abelian models in the one-dimensional and the two-dimensional cases.
Finally, in Sec.~\ref{sec:4} we discuss our results and possible experimental realizations.

\section{wavepacket dynamics of motion}
\label{sec:2}

In this section, we discuss the semiclassical EOM of a wavepacket in a periodic potential under an external inhomogeneous electric field. 
The single-particle Hamiltonian (with units $\hbar = 1$) is
\begin{equation}
\hat{H} = \hat{H}_0 + \hat{V}_{\mathrm{ext}} \,,
\end{equation}
where $\hat{H}_0 = \hat{\mathbf p}^2/2m + \hat{V}(\mathbf{r})$ is the Hamiltonian of a particle with mass $m$ in a periodic potential $\hat{V}(\mathbf{r})$ and 
the external potential is given by 
\begin{equation}
\hat{V}_{\mathrm{ext}} = E^\mu \hat{r}_\mu + \frac{1}{2} E^{\mu\nu} \hat{r}_\mu \hat{r}_\nu + \frac{1}{6} E^{\mu\nu\rho} \hat{r}_\mu \hat{r}_\nu \hat{r}_\rho\,,
\end{equation}
and includes linear, quadratic, and cubic terms only. 
Here, $\hat{r}_\mu$ and $\hat{p}^\nu$ are the position and momentum operators for a single particle in spatial dimensions $D$, which satisfy the commutation relation $[\hat{r}_\mu,\hat{p}^\nu] = i\delta_{\mu}^{\nu}$. 
The periodic structure of $\hat{V}(\mathbf{r})$ allows us to describe the system using the Bloch states $\ket{\psi^{(n)}_{\mathbf{q}}} = e^{i \mathbf{q}\cdot \mathbf{r}} \ket{u^{(n)}_{\mathbf{q}}} $ for the $n$-th band, which are the eigenstates of $\hat{H}_0$, and ${\mathbf{q}}$ is the quasimomentum. 
The function $u^{(n)}_{\mathbf{q}}(\mathbf{r}) := \braket{\mathbf{r}|u^{(n)}_{\mathbf{q}}}$ has the same periodicity as the crystal lattice and the Bloch states satisfy $\hat{H}_0 \ket{\psi^{(n)}_{\mathbf{q}}} = \mathcal{E}_n(\mathbf{q})\ket{\psi^{(n)}_{\mathbf{q}}}$, with $\mathcal{E}_n(\mathbf{q})$ the band dispersion.
In what follows, we will discuss the dynamics of a generic wavepacket formed by Bloch states of multiple bands (non-Abelian case) that can be represented as $\ket{\Psi} = \mathcal{N}\sum_{n,\mathbf{q}} c_n(\mathbf{q}) \ket{\psi_{\mathbf{q}}^{(n)}}$ where $\mathcal{N}$ is a normalization coefficient.
Here below, we review the EOM that we are discussing in this work.

\subsection{Wavepacket trajectory}

For a wavepacket $\ket{\Psi}$ in the form indicated above, the mean projected position $\mathbf{R}=\braket{\hat{\mathbf{r}}}_{\mathcal{S}}$ satisfies:
\begin{widetext}
\begin{equation}
\begin{split}
\dot{R}_\mu &= \left\langle [\hat{\boldsymbol{D}}_\mu, \hat{H}_0] \right\rangle_{\mathcal{S}} + E^\nu(\mathbf{R}) \left\langle \hat{\boldsymbol{\Omega}}_{\mu\nu} \right\rangle_{\mathcal{S}} + \frac{1}{2}E^{\nu\rho}(\mathbf{R}) \left\langle \left[ \hat{\boldsymbol{D}}_\mu, \hat{\boldsymbol{g}}_{\nu\rho} \right] \right\rangle_{\mathcal{S}} \cr
& + \frac{1}{6} E^{\nu\rho\lambda} \left[ \frac{1}{2} \left\langle \left\{ \hat{\boldsymbol{\Omega}}_{\mu\nu}, \hat{\boldsymbol{g}}_{\rho\lambda} \right\} \right\rangle_{\mathcal{S}} + (\nu,\rho,\lambda)_P \right] 
+ \frac{1}{6} E^{\nu\rho\lambda} \left\langle \left[ \hat{\boldsymbol{D}}_\mu, \hat{\boldsymbol{T}}_{\nu\rho\lambda} \right] \right\rangle_{\mathcal{S}} + \delta\dot R_\mu \,,
\end{split}
\label{eq:R}
\end{equation}
\end{widetext}
where $\hat{\boldsymbol{D}}_\mu = \partial_\mu - i \hat{\mathbf{A}}_\mu$ is the gauge covariant derivative with $\partial_\mu \equiv \partial/\partial q^\mu$ and $A_\mu^{(m,n)} \equiv i \langle u_\mathbf q^{(m)}| \partial_\mu u_\mathbf q^{(n)}\rangle$. 
The symbol $\braket{\hat{O}}_\mathcal{S}$ corresponds to $\braket{\hat{O}}_\mathcal{S} \equiv \braket{ \hat{P}_{\mathcal{S}} \hat{O} \hat{P}_{\mathcal{S}}}$, where $\mathcal{S}$ represents a subspace of bands and its projection operator is $\hat{P}_{\mathcal{S}}$. 
The notation $(\nu,\rho,\lambda)_P$ represents a cyclic permutation of the three dummy indices $\nu,\rho,\lambda$. 
The functions $E^\nu (\mathbf{R})$ and $E^{\nu\rho}(\mathbf{R})$ for a wavepacket with mean position $\mathbf{R}$ have the form $E^\nu(\mathbf R) = E^\nu + (E^{\nu\nu} R_\nu + E^{\nu\rho} R_{\rho})/2 + ( E^{\nu\rho\lambda} R_\rho R_\lambda + 2 E^{\nu\nu\rho} R_\nu R_\rho )/6$ and $E^{\nu\rho}(\mathbf{R}) = E^{\nu\rho} + ( E^{\nu\nu\rho} R_\nu + E^{\nu\rho\rho} R_\rho + E^{\nu\rho\lambda} R_\lambda)/6$.
The non-Abelian Berry curvature and quantum metric tensors~\cite{Mayuquan-PhysRevB.81.245129,PhysRevA.109.043305} are respectively given by
\begin{equation}
\hat{\boldsymbol{\Omega}}_{\mu\nu} = \partial_\mu \hat{\mathbf{A}}_\nu - \partial_\nu \hat{\mathbf{A}}_\mu -i [\hat{\mathbf{A}}_\mu, \hat{\mathbf{A}}_\nu]\,,
\end{equation}
and
\begin{equation}
\begin{split}
g^{(m,n)}_{\mu\nu} &= \frac{1}{2} \braket{ \partial_\mu u^{(m)}| \partial_\nu u^{(n)}} - \frac{1}{2}\sum_{l\in \mathcal{S}} A^{(m,l)}_{\mu} A^{(l,n)}_{\nu} \cr
&+ (\mu\leftrightarrow\nu) \,.
\end{split}
\label{eq:metric}
\end{equation}
We note that the velocity contribution induced by the effective quadratic potential in Eq.~\eqref{eq:R} (i.e. the term proportional to $E^{\nu\rho}(\mathbf R)$) includes terms related to the derivative of the quantum metric, as first proposed in Ref.~\cite{Lapa-PhysRevB.99.121111}.
In addition, we find that the non-Abelian case also brings a commutator $[\mathbf{A}_\mu, \boldsymbol g_{\nu\rho}]$ that appears within the covariant derivative.
In Eq.~\eqref{eq:R}, the cubic potential contributions ($E^{\nu\rho\lambda}$) yields two terms, whose Abelian form was introduced in Ref.~\cite{Kozii-PhysRevLett.126.156602} and that we generalize to the non-Abelian case. 
The first term contains the product of the quantum metric and the Berry curvature, which leads to a geometric current.
The second term contains a rank-3 tensor $\hat{\boldsymbol{T}}_{\nu\rho\lambda}$ that,  for the Abelian case Ref.~\cite{Kozii-PhysRevLett.126.156602} is identified as the gauge-invariant part of the third cumulant of the position operator $\braket{\delta \hat r_\nu \delta \hat r_\rho \delta \hat r_\lambda}_\mathcal{S}$ and $\delta \hat{r} = \hat{r} - \braket{\hat{r}}_\mathcal{S}$. Notice that this tensor is also mathematically related to a quantum geometric connection \cite{Ahn-PhysRevX.10.041041}. 
In this work, we find the generalization of this tensor to the non-Abelian case and we provide its definition and additional details in Supplemental Material due to its complicated form.
Furthermore, there also exist shape-dependent terms, $ \delta\dot R_\mu$, whose extended form is provided in the Supplemental Material and can be neglected when the wavepacket is narrow in real space \cite{Kozii-PhysRevLett.126.156602}.
Such terms, that also appear in the equations for other quantities discussed below, depend non-trivially on the wavepacket shape represented by the coefficients $c_n(\mathbf{q})$.

For the mean momentum of the wavepacket, $\mathbf{Q} = \braket{\hat{\mathbf{q}}}_\mathcal{S}$, the cubic term of the potential yields a contribution in addition to the standard linear term that reads 
\begin{align}
\dot{Q}^\mu &= - E^\mu(\mathbf{R}) - \frac{1}{2} E^{\mu\nu\rho} W_{R,\nu\rho} \,. \label{eq:Q}
\end{align}
The first term corresponds to Newton's law $\dot{Q}^\mu = -\partial_\mu V_{\mathrm{ext}}(\mathbf{r}) = -E^{\nu}(\mathbf{R})$. 
The second term of Eq.~\eqref{eq:Q} contains the wavepacket variance \mbox{$W_{R,\mu\nu} = \braket{\hat{r}_\mu \hat{r}_\nu }_{\mathcal{S}} - \braket{\hat{r}_\mu}_{\mathcal{S}} \braket{\hat{r}_\nu}_{\mathcal{S}}$}.
In the Supplemental Material we show that the variance splits into $W_{R, \mu\nu} = \braket{\boldsymbol{g}_{\mu\nu}}_\mathcal{S} + \dots$, where the dots indicate a shape-dependent contribution. 
This form is the one displayed in Ref.~\cite{Kozii-PhysRevLett.126.156602}; however, to be best of our knowledge the simplified expression presented in Eq.~\eqref{eq:Q} in terms of the variance is a genuine result of this work.
We have noticed that in practical numerical simulations, this equation remains valid only as long as the momentum-space wavepacket width has not grown to the extent that it wraps around the entire Brillouin zone.
We will comment further about this point later.

\subsection{Wavepacket variance}

In addition to studying the wavepacket's trajectory, we also investigate the equations governing its shape in both real and momentum space. 
Here we present the equations for the variance under a quadratic potential, whereas we leave the discussion of the more involved cubic terms to future work.
The real space variance of the wavepacket, $W_{R,\mu\nu}$, satisfies
\begin{equation}
\begin{split}
\dot{W}_{R,\mu\nu} &= - i\left\langle \left[ \hat{\boldsymbol{g}}_{\mu\nu}, \hat H_0 \right] \right\rangle_{\mathcal{S}} - E^\rho(\mathbf{R}) \left\langle \left[ \hat{\boldsymbol{D}}_\rho, \hat{\boldsymbol{g}}_{\mu\nu} \right] \right\rangle_{\mathcal{S}} \cr
&- \frac{i}{2} E^{\rho\lambda} \left\langle \left[ \hat{\boldsymbol{g}}_{\mu\nu}, \hat{\boldsymbol{g}}_{\rho\lambda} \right] \right\rangle_{\mathcal{S}} + \delta \dot{W}_{R,\mu\nu} \,. \cr
\end{split}
\label{eq:VarR}
\end{equation}
The Abelian version of this equation in the presence of linear electric fields in one-dimensional systems was proposed in Ref.~\cite{Souza-PhysRevB.79.045127}, whereas here we generalize it to higher-dimensional non-Abelian systems up to quadratic potentials. 
The first term on the right-hand side of Eq.~\eqref{eq:VarR} is relevant in multi-band cases unless the bands are perfectly degenerate and for which it vanishes.
The second term shows a dependence on the effective linear potential $E^\rho(\mathbf R)$ and it shows that the real space variance depends on the metric dipole \cite{Souza-PhysRevB.79.045127} as well. 
Importantly, we find that a quadratic potential only affects the variance dynamics in dimensions $D\ge 2$ and in the multi-band non-Abelian case via the commutator $[\hat{\boldsymbol{g}}_{\mu\nu}, \hat{\boldsymbol{g}}_{\rho\lambda}]$.
The numerical validation of this term is not performed in this work but it is left to future investigation. Finally, we also identify other shape-dependent terms, $\delta \dot{W}_{R,\mu\nu}$,  provided in the Supplemental Material.

Differently from the real space variance, where a linear potential already affects its evolution, we have found that an inhomogeneous electric field is uniquely responsible for the evolution of the wavepacket's momentum variance $W_{Q}^{\mu\nu} = \braket{  \hat{q}^\mu \hat{q}^\nu  }_{\mathcal{S}} - \braket{\hat{q}^\mu}_{\mathcal{S}} \braket{\hat{q}^\nu}_{\mathcal{S}} $. 
Actually, the presence of an inhomogeneous potential causes the wavepacket expand across the entire Brillouin zone, ultimately forming an interference pattern. 
The dynamics of the momentum space variance under a quadratic potential can be described by the following equation 
\begin{equation}
\dot{W}_{Q}^{\mu\nu} =  -E^{\mu\rho} W_{RQ,\rho}^{\nu} + ( \mu \leftrightarrow \nu) \,.
\label{eq:WQ}
\end{equation}
This equation contains a phase space variance that reads $W_{RQ,\mu}^{\nu} = \braket{\hat{r}_\mu \hat{q}^\nu  }_{\mathcal{S}} - \braket{\hat{r}_\mu}_{\mathcal{S}} \braket{\hat{q}^\nu}_{\mathcal{S}} $.
While the term $\braket{\hat{r}_\mu}_{\mathcal{S}} \braket{\hat{q}^\nu}_{\mathcal{S}}$ can be readily obtained via solving the Eq.~\eqref{eq:R} and Eq.~\eqref{eq:Q}, we find the appearance of the term $\braket{ \hat{r}_\mu\hat{q}^\nu }_{\mathcal{S}}$, which requires a separate treatment. 
The corresponding equations read
\begin{equation}
\dot{W}_{RQ,\mu}^\nu = - E^{\nu\rho} W_{R,\mu\rho} + \delta \dot{W}_{RQ,\mu}^\nu  \,, 
\label{eq:VarRQ}
\end{equation}
where we see that the phase space variance relates to the real space variance up to shape-dependent terms.
In what follows, we will consider wavepackets with large space variance such that the first term dominates over the second and can be disregarded with a good approximation.

So far, we have developed a set of semiclassical equations that describe the behavior of a wavepacket under inhomogeneous external potentials and we have related them to the quantum geometry. 
In the following section, we numerically examine the equations that we have presented and introduced in this section.
Notice that in the formulation above, we do not provide the equations for $\dot c_n(\mathbf q)$, see for example Ref.~\cite{Niu-RevModPhys.82.1959}, which rule the band occupation dynamics. 
Differently from the well-known case of homogeneous electric fields, where the solution is given by the path-ordered Wilson loop, in the presence of inhomogeneous fields these equations become non-local in momentum space and cannot be straightforwardly solved (not shown here).
Further investigation of these equations could shed interesting results on the non-Abelian band dynamics.

\section{Numerical analysis of EOM}
\label{sec:3}

In this section, we focus on validating and analyzing Eqs.~\eqref{eq:R},~\eqref{eq:Q}~\eqref{eq:WQ} and~\eqref{eq:VarRQ} in some simple models where we identify \emph{i)} \emph{geometrically induced Bloch oscillations} in the wavepacket displacement and variance and \emph{ii) transverse displacement in the absence of Berry curvature}. 
In particular, we select models with flat bands, where the band dispersion plays no role. 
Transport or dynamical properties are thus solely originating by the interplay of inhomogeneous electric fields and band geometry. 
Notice that dynamical effects originating from quadratic or cubic electric fields have been less explored, as compared to those occurring in the presence of linear ones. 
For example, the latter have been extensively studied in previous works for cold atoms in optical lattices, especially with a focus on the effects of the Berry curvature and anomalous velocity~\cite{Price-PhysRevA.85.033620,Goldman-aidelsburger2015measuring,Goldman_2014,Goldman-2016Nat.Phys,Goldman-PhysRevLett.111.135302,Marco-ncs41467-020-19518-x}. 

\begin{figure}[!b]
\centering
\includegraphics[width=8.5cm]{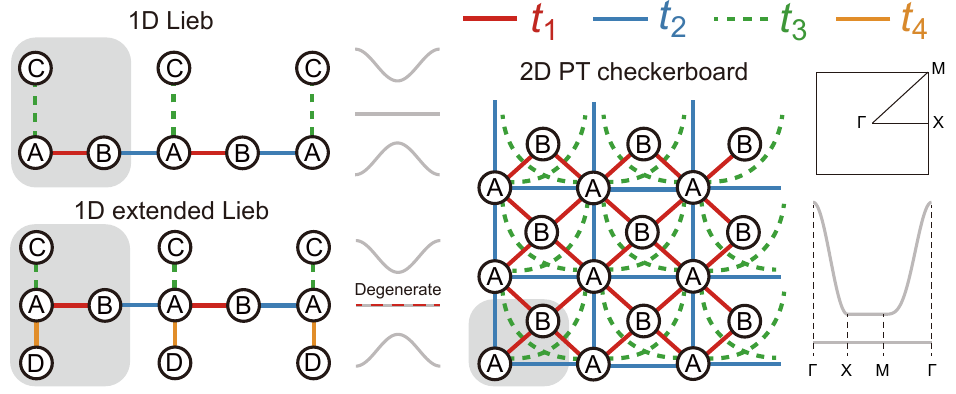}
\caption{Tight-binding models. One dimensional Lieb model, extended Lieb model and two dimensional checkerboard model's hopping structure and band spectrum with single and degenerate flat bands. Corresponding parameters are those used in the plots of the next figures. The extended Lieb lattice model is discussed more in detail in the Supplemental Material.
}
\label{fig:Models}
\end{figure}

Specifically, in one dimension, we study Lieb lattice models, while
in two dimensions, we study a checkerboard model with $\mathcal{PT}$ symmetry thus displaying no Berry curvature. 
The tight-binding structure of these models is illustrated in Fig.~\ref{fig:Models}.

\subsection{One dimensional Lieb model}

The one-dimensional Lieb model ~\cite{Baboux-PhysRevLett.116.066402,Biondi-PhysRevLett.115.143601,Hyrkas-PhysRevA.87.023614} considered in Fig.~\ref{fig:Models} is given by
\begin{equation}
H_{1} = \sum_j (t_1 b^\dag_j a_j + t_2 a_{j+1}^\dag b_j + t_3 c_j^\dag a_j) + \text{H.c.}\,,
\end{equation}
with Bloch Hamiltonian 
\begin{equation}
\mathcal{H}_1(q^x) = \left( \begin{matrix}
0 & t_1 + t_2 e^{-i q^x a} & t_3 \\
t_1 + t_2 e^{i q^x a} & 0 & 0 \\
t_3 & 0 & 0 \\
\end{matrix} \right) \,,
\end{equation}
with $a$ the lattice spacing.
Due to chiral symmetry the model has a protected zero energy band whose eigenstate reads $\ket{u_{\mathrm{FB}}(q^x)} = \mathcal{N}(0, -t_3, t_1 \!+\! t_2 e^{-iq^x a})^T$, where $\mathcal{N}$ is a normalization factor.
The corresponding gap from either of the dispersive bands is given by $\Delta_{1}(q^x) = \sqrt{t_1^2 + t_2^2 + t_3^2 + 2 t_1 t_2 \cos{(q^x a)}}$ and requires $t_3\neq 0$ to be open for all momenta. The quantum metric $g_{xx}$ can be derived from Eq.~\eqref{eq:metric} and reads
\begin{align}
g_{xx}(q^x) &= \frac{t_2^2 t_3^2}{[\Delta_1(q^x)]^4} \,.
\end{align}
The rank-3 gauge-invariant tensor $T_{xxx}$ is defined in the Supplemental Material and, for the one-dimensional Abelian case treated here, reads
\begin{equation}
\begin{split}
T_{xxx}(q^x) &= \partial^2_xA_x - 3 A_{x}(g_{xx} + \partial_x A_x) - A_x^3 \cr
&- i\frac{3}{2} \partial_x g_{xx} - i \braket{u_{\mathrm{FB}}|\partial^3_x u_{\mathrm{FB}}}\,,
\end{split}
\end{equation}
which, for the model analyzed in this section takes the form
\begin{align}
T_{xxx}(q^x) &= \frac{t_2^2 t_3^2(t_1^2-t_2^2+t_3^2)}{[\Delta_1(q^x)]^6} \,. \label{eq:1D Lieb rank3}
\end{align}

\subsubsection{Dynamics with linear and quadratic potentials}

We thus consider the dynamics of a wavepacket on the flat band subjects to external linear and quadratic electric potentials of the form 
\begin{equation}
 V_{\mathrm{ext}} = E^x x + \frac{1}{2}E^{xx} x^2\,.
\end{equation}
From Eqs.~\eqref{eq:R},~\eqref{eq:Q},~\eqref{eq:WQ} and ~\eqref{eq:VarRQ}, we find that the dynamics is described by 
\begin{align}
\dot{R}_x &= \frac{1}{2} E^{xx} \left\langle\frac{\partial g_{xx}}{\partial q^x} \right\rangle_{\mathcal{S}} \,,
\label{eq:1d Lieb R} \\
\dot{Q}^x &= -E^x(R_x) \,,
\label{eq:1d Lieb Q} \\
\dot{W}_{R,xx} &= -E^x(R_x) \left\langle \frac{\partial g_{xx}}{\partial q^x}\right\rangle_{\mathcal{S}} \,,
\label{eq:1d Lieb VarR} \\
\dot{W}_{Q}^{xx} &= -2 E^{xx} W_{RQ,x}^{x} \,,
\label{eq:1d Lieb dVarQ} \\
\dot{W}_{RQ,x}^{x} &= - E^{xx} W_{R,xx} \,,
\label{eq:1d Lieb dVarRQ}
\end{align}
where $E^{x}(R_x) = E^x + E^{xx} R_x$. 
In the equations above, all shape-dependent terms are either zero for a flat band model in one dimension or negligible for the purpose of the simulations performed here, and are therefore discarded.
We now assume that the wavepacket remains Gaussian over time, thus the occupation probabilities, $|c(q_x,t)|^2$, read
\begin{equation}
|c(q^x,t)|^2 \approx \mathcal{N}\exp{\left[ - \frac{(q^x - Q^x(t))^2}{2 W_{Q}^{xx}(t)} \right]}\,,
\label{eq:1D Lieb c(q,t)}
\end{equation}
where $\mathcal{N}$ is a normalization coefficient. 
We can solve for $W_{Q}^{xx}(t)$ by employing Eq.~\eqref{eq:1d Lieb dVarQ} and Eq.~\eqref{eq:1d Lieb dVarRQ} and assuming that $W_{R,xx}(t)\approx W_{R,xx}(0)$, namely that the dynamics of the real space variance is not affecting the dynamics of the momentum space variance. 
These considerations yield
\begin{equation}
\begin{split}
\Delta W_{Q}^{xx}(t) &= 2 E^{xx} W_{RQ,x}^{x}(0) t + (E^{xx})^2 W_{R,xx}(0) t^2 \,, 
\label{eq:1d Lieb ΔVarQ}
\end{split}
\end{equation}
where $\Delta W_{Q}^{xx}(t) = W_{Q}^{xx}(t) - W_{Q}^{xx}(0)$.
We have now established a set of equations that are closed and determine the main dynamical properties to be compared with the numerical results.

\begin{figure}[!t]
\centering
\includegraphics[width=8.5cm]{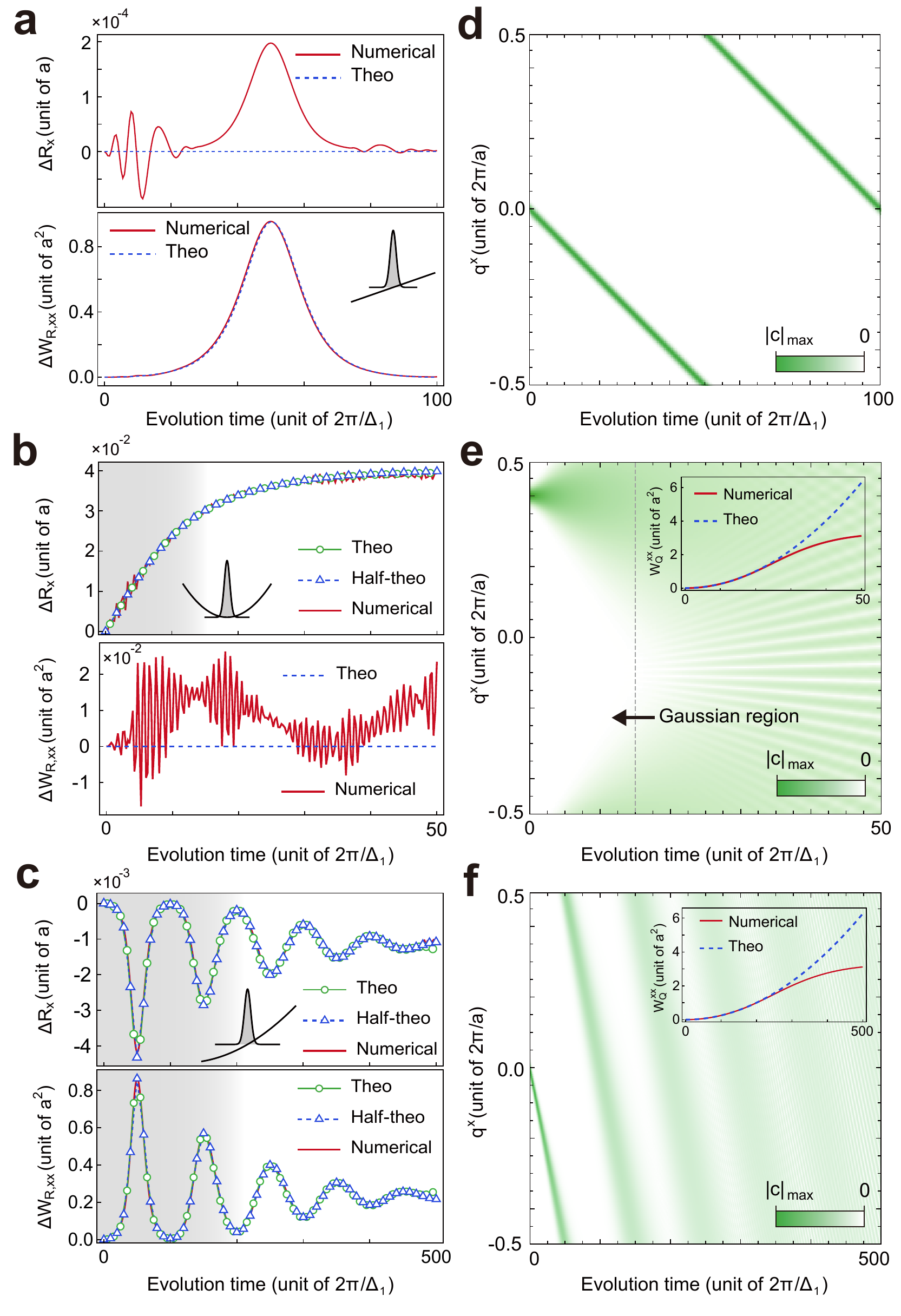}
\caption{ Wavepacket dynamics in 1D Lieb lattice with linear and quadratic potentials. Evolution of $\Delta R_x$ and $\Delta W_{R,xx}$ under an external (a) linear potential with strength $E^x/\Delta_{1} = 10^{-2}$, (b) quadratic potential with strength $E^{xx}/\Delta_{1} = 10^{-3}$, and (c) a combination of both, $(E^x,E^{xx})/\Delta_{1} = (10^{-2},10^{-4})$. 
For the theoretical curves we use the semiclassical equations assuming the population remains Gaussian over time. For the half-theoretical curves we input the population of the different states, $|c(q_x,t)|^2$, from the numerical results.
The corresponding momentum space distributions are shown in panels (d-f) and their insets show that the momentum space variance evolves as $W_{Q}^{xx}\sim t^2$ at short time, as indicated in the main text. Other parameters are set as $t_1 = t_2 = t_3 = \Delta_{1}$ in a finite system size of length $L=200$ sites. The wave packet is initialized with a momentum space standard deviation of $\sigma_Q=0.02\pi/a$ (the real space standard deviation is $\sigma_R = 7.99a$ and $ 7.96 a$ for $Q=0$ and $Q=0.8\pi/a$, respectively). 
}
\label{fig:1d Lieb quadratic EOMs}
\end{figure}

In Fig.~\ref{fig:1d Lieb quadratic EOMs},
we plot the wavepacket's trajectories under different external potentials. 
In the case of a linear potential ($E^{xx} = 0$), the wavepacket's velocity remains close to zero, but the evolution of its variance reveals a Bloch oscillation that reproduces the shape of the quantum metric distribution, and follows Eq.~\eqref{eq:1d Lieb VarR}, as shown in Fig.~\ref{fig:1d Lieb quadratic EOMs}(a). 
The small real space displacement shown in the upper panel of Fig.~\ref{fig:1d Lieb quadratic EOMs}(a) originates from non-adiabatic effects and is not captured by our projected equations.
The momentum space wavepacket distribution follows Eq.~\eqref{eq:1d Lieb Q} which is plotted in Fig.~\ref{fig:1d Lieb quadratic EOMs}(d).

Under a quadratic potential ($E^x = 0$), the velocity obeys Eq.~\eqref{eq:1d Lieb R} and the displacement is caused by the quantum metric dipole \cite{Xiao}, $\braket{\partial g_{xx}/\partial q^x}_{\mathcal{S}}$.
To observe a noticeable displacement, we set the initial momentum to $Q^x(0) = 0.8\pi/a$, ensuring that $\braket{\partial g_{xx}/\partial q^x}_{\mathcal{S}}$ is as large as possible. 
In Fig.~\ref{fig:1d Lieb quadratic EOMs}(b), we compare the purely numerical results for the trajectory evolution (red solid line) with those obtained by the semiclassical equations calculated using Eq.~\eqref{eq:1d Lieb R} (blue dashed line with triangle markers). 
The quadratic potential causes the  momentum space distribution, $|c(q_x,t)|^2$, to change in time, thus generating a variation of the momentum space variance according to Eqs.~\eqref{eq:1d Lieb dVarQ}, \eqref{eq:1d Lieb dVarRQ}, which we simplified into Eq.~\eqref{eq:1d Lieb ΔVarQ}. 
We find that the Gaussian approximation for $|c(q_x,t)|^2$, Eq.~\eqref{eq:1D Lieb c(q,t)}, provides a good description of the dynamics, as shown by the green solid line with circle markers in Fig.~\ref{fig:1d Lieb quadratic EOMs}(b).
Notice that no real-space Bloch oscillations take place, as the average momentum distribution $Q^x$ remains constant.  
However, in our numerics we identify a very small non-vanishing dynamics for the real space variance which originates from non-adiabatic effects, see lower panel in Fig.~\ref{fig:1d Lieb quadratic EOMs}(b). 
Moreover, we also extract the momentum distribution, $|c(q_x,t)|^2$, from Fourier transforming the exact numerical evolution of the wavepacket. 
As shown in Fig.~\ref{fig:1d Lieb quadratic EOMs}(e), the evolution of the momentum distribution is symmetric around the initial momentum, reflecting $\dot{Q}^{x} \approx 0$. 

When the linear and quadratic potential terms are both present, we find the appearance of a (damped) Bloch oscillation occurring both for the real space position and variance, as shown in Fig.~\ref{fig:1d Lieb quadratic EOMs}(c). 
We stress that this effect requires the simultaneous presence of the linear and quadratic terms of the external potential, as one can deduce from inspecting Eq.~\eqref{eq:1d Lieb R} and Eq.~\eqref{eq:1d Lieb Q}.

Here we conclude that we can predict the wavepacket's evolution at short times by only using its initial macroscopic properties, \emph{i.e.} the wavepacket's position $R_x(0)$, momentum $Q^x(0)$ and variances $W_{R,xx}(0),~W_{Q}^{xx}(0)$. 
The gray regions in Fig.~\ref{fig:1d Lieb quadratic EOMs}(b,c) indicate the timescale during which Eq.~\eqref{eq:1D Lieb c(q,t)} is valid, namely when the momentum space distribution remains Gaussian. 
This is reflected in the short-time agreement between the numerically evaluated $W_{Q}^{xx}$ and our theoretical prediction from Eq.~\eqref{eq:1d Lieb ΔVarQ}, as shown in the inset of Fig.~\ref{fig:1d Lieb quadratic EOMs}(e). 
Notice that the variance evolves with $t^2$, thus showing that the linear term $\sim t$ of Eq.~\eqref{eq:1d Lieb ΔVarQ} can be neglected. 
Additional discussions about this point are detailed in the Supplemental Material.
At later times, the Gaussian approximation breaks down and leads to a significant offset between the theoretical trajectory and the numerically exact one.

\begin{figure}[!t]
\centering
\includegraphics[width=8.5cm]{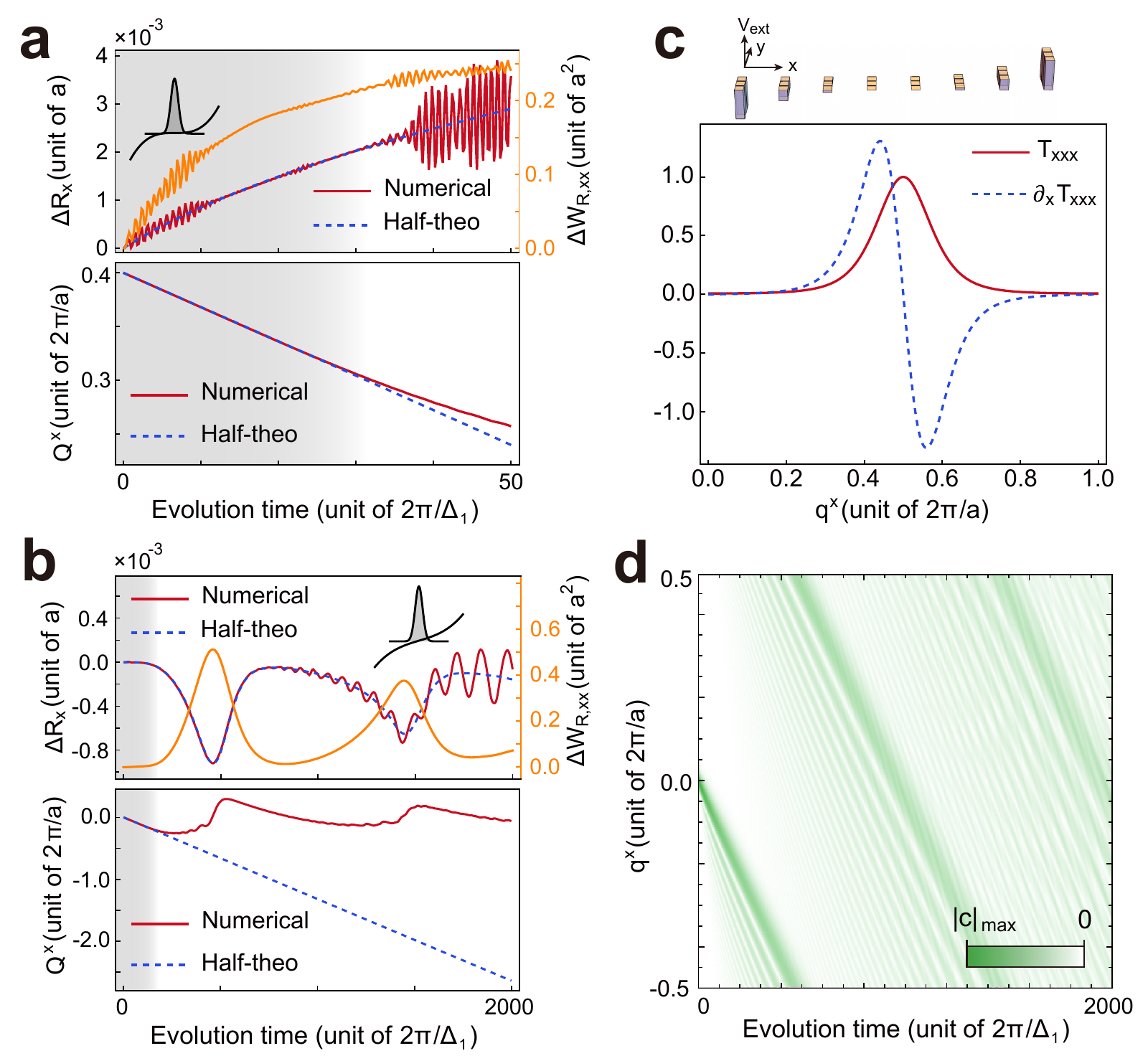}
\caption{ Wavepacket dynamics in 1D Lieb lattice with linear and cubic potentials. Wavepacket trajectory and variance in real and momentum space with potential strength (a) $E^{xxx}/\Delta_1 = 10^{-4}$ and (b) $(E^{x}, E^{xxx})/\Delta_1 = (10^{-3},10^{-5})$. (c) External cubic potential, rank-3 gauge invariant tensor $T_{xxx}$ and its dipole moment. (d) Corresponding distribution of $|c(q_x,t)|$ under weak linear and cubic external potential. Other parameters are the same as in Fig.~\ref{fig:1d Lieb quadratic EOMs}. }
\label{fig:1D Lieb cubic EOMs}
\end{figure}

\subsubsection{Dynamics with a cubic potential}

We now discuss the dynamics in the presence of an external cubic potential
\begin{equation}
V_{\mathrm{ext}} = E^{x}x +\frac{1}{6}E^{xxx} x^3\,,
\end{equation}
where we also included the linear term to observe nontrivial dynamics.
The wavepacket dynamics is ruled by
\begin{align}
\dot{R}_x &= \frac{1}{6} E^{xxx} \left\langle \frac{\partial T_{xxx}}{\partial q^x} \right\rangle_{\mathcal{S}} \,, \label{eq:1d Lieb cubic dR} \\
\dot{Q}^x &= -E^x(R_x) - \frac{1}{2} E^{xxx}  W_{R,xx}  \,,
\label{eq:1d Lieb cubic dQ}
\end{align}
where $E^{x}(R_x) = E^x + \frac{1}{2}E^{xxx}R_x^2$.
Notice that we have derived the equations for the variance, Eq.~\eqref{eq:VarR}, up to quadratic potentials. 
Therefore, in order to solve the equations of motion given above we will rely on a numerical estimation of the variance dynamics, differently from what we discussed in the previous section.
Finally, it is worth noticing that the effects originating from cubic potentials are significantly smaller than those coming from quadratic and linear ones. 
Thus, identifying the corresponding dynamics is harder, as noise originating from non-adiabatic effects becomes more significant: a tiny population in other bands, i.e. a non-adiabatic contribution, may overwhelm the signal.

In Fig.~\ref{fig:1D Lieb cubic EOMs}(c), we plot the momentum dependence of the rank-3 tensor $T_{xxx}$ and of its dipole moment $\partial T_{xxx}/\partial q^x$ from Eq.~\eqref{eq:1D Lieb rank3}, which shows a shape analogous to the quantum metric. 
In Fig.~\ref{fig:1D Lieb cubic EOMs}(a), we display the evolution trajectories of wavepackets with initial momentum chosen to maximize $|\partial T_{xxx}/\partial q_x|$ and thus the initial velocity of the wavepacket. 
We find the theoretical prediction calculated using Eq.~\eqref{eq:1d Lieb cubic dR}-\eqref{eq:1d Lieb cubic dQ} consistent with the numerical results, and the high-frequency and high-amplitude oscillations arise from non-adiabatic effects. 
Moreover, we observe that under a cubic potential the distribution $|c(q_x,t)|^2$ no longer remains Gaussian over the time evolution, as shown in Fig.~\ref{fig:1D Lieb cubic EOMs}(d).
For this reason, to compute the theoretical curves we resort to the numerical estimate of the occupations $|c(q_x,t)|^2$ that we input in the semiclassical equations to obtain the theory curves in Fig.~\ref{fig:1D Lieb cubic EOMs}(a,b).
We also analyze the interplay of linear and cubic potentials, where we take the linear term sufficiently weak. 
Similarly to the previous section, see Fig.~\ref{fig:1d Lieb quadratic EOMs}(c), we find geometric Bloch oscillations induced by the rank-3 tensor $T_{xxx}$, which is shown in Fig.~\ref{fig:1D Lieb cubic EOMs}(b).

\begin{figure*}[!t]
\centering
\includegraphics[width=17cm]{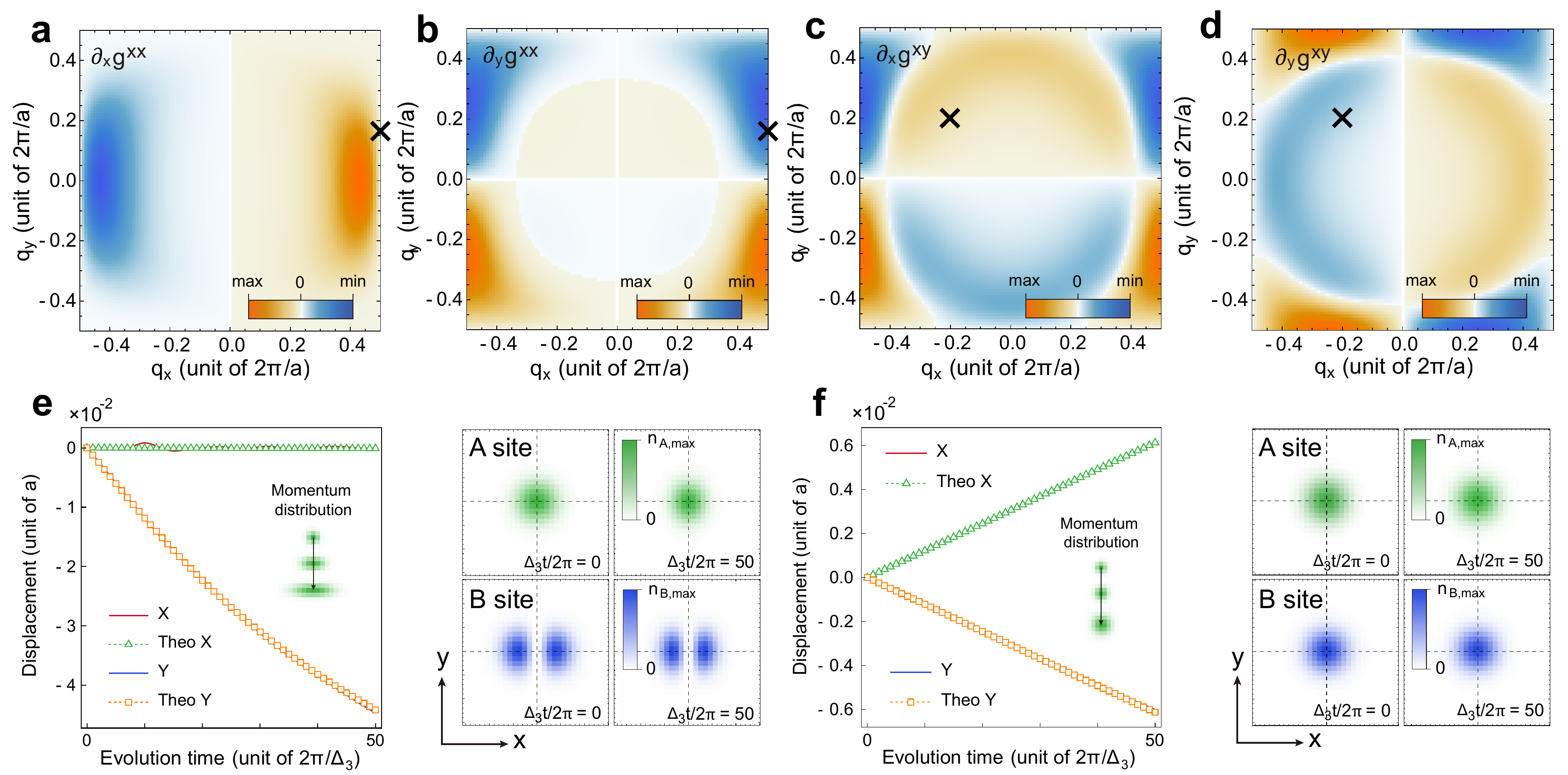}
\caption{Wavepacket dynamics in the 2D checkerboard lattice. (a-d) Distributions of quantum metric dipole for the flat band. (e-f) Trajectory of the wavepacket under a quadratic potential and sublattice occupation. 
In (e), we consider an initial momentum  $(Q_x, Q_y) = (\pi/a, \pi/3a)$ and quadratic potential with strength $E^{xx}/\Delta_2 = 10^{-3}$. 
In (f) we take initial momentum at $(Q_x, Q_y)=(-2\pi/5a, 2\pi/5a)$ and  $E^{xy}/\Delta_2 = 10^{-3}$. 
Other parameters are $t_1/\Delta_2 = -t_2/\Delta_2  = -2t_3/\Delta_2 = -1/2$ and $\delta_0/\Delta_2  = -1$, $\sigma_Q = 0.06\pi/a$.}
\label{fig:2D checkerboard}
\end{figure*}

\subsection{\texorpdfstring{Two dimensional $\mathcal{PT}$-invariant checkerboard model}{Two dimensional PT-invariant checkerboard model}}

Here, we consider a $\mathcal{PT}$-invariant checkerboard model, such that the Bloch Hamiltonian satisfies $ \mathcal H^\ast(\mathbf{q}) = \mathcal H(\mathbf{q})$ thus displaying real eigenvectors $\ket{u(\mathbf{q})}\in \mathbb R^N$.
In this way the Berry curvature is identically zero and the Hall displacement is absent based on the semiclassical EOM.
We can thus focus only on geometric effects originating from a quadratic potential. 
The two-dimensional $\mathcal{PT}$-invariant checkerboard model discussed here is shown in Fig.~\ref{fig:Models} and has two sublattices, $A$ and $B$. It is related to the one introduced in Ref.~\cite{Ahn-PhysRevX.9.021013} and is described by the following tight-binding Hamiltonian
\begin{equation}
H_{3} = H_{nn} + H_{nnn} + H_{nnnn} + \delta_0 \sum_{j,l} b_{j,l}^\dag b_{j,l} \,,
\end{equation}
where we defined the hopping terms as
\begin{equation}
\begin{split}
H_{nn} &= t_1 \sum_{j,l} \left( b_{j,l}^\dag a_{j,l} + b_{j+1,l}^\dag a_{j,l} + b_{j,l+1}^\dag a_{j,l} \right. \cr
& \left.+ b_{j+1,l+1}^\dag a_{j,l} \right) + \mathrm{H.c.} \,, \cr
H_{nnn} &= t_2 \sum_{j,l} ( a^\dag_{j+1,l} a_{j,l} + a^\dag_{j,l+1} a_{j,l} ) + \mathrm{H.c.} \,, \cr
H_{nnnn} &= t_3 \sum_{j,l} (a_{j+1,l+1}^\dag a_{j,l} + a_{j-1,l+1}^\dag a_{j,l}) + \mathrm{H.c.}
\end{split}
\end{equation}
and we introduce an onsite energy $\delta_0$ for the $B$ sites.
The corresponding Bloch Hamiltonian reads $\mathcal{H}_3(\mathbf{q}) = d_x(\mathbf{q}) \sigma_x + d_z(\mathbf{q}) \sigma_z + d_0(\mathbf{q})$ where
\begin{equation}
\begin{split}
d_x(\mathbf{q}) &= 4 t_1 \cos({\frac{q^x a}{2})} \cos{(\frac{q^y a}{2})} \,, \cr
d_z(\mathbf{q}) &= - \delta_0/2 +t_2 \left[ \cos{(q^x a)} + \cos{(q^y a)} \right] \cr
&+ t_3 \left[ \cos{(q^x a + q^y a) } + \cos{(q^x a - q^y a)} \right] \,, \cr
d_0(\mathbf{q}) &= \delta_0/2  + t_2 \left[ \cos{(q^x a)} + \cos{(q^y a)} \right] \cr
&+ t_3 \left[ \cos{(q^x a + q^y a) } + \cos{(q^x a - q^y a)} \right] \,. \cr
\end{split}
\end{equation}
To recover a flat band, we set the parameters as $t_2 = 2 t_3$, $\delta_0 = t_1^2/t_3 - 4 t_3$. The corresponding minimal gap is $\Delta_{2} = |t_1^2/t_3|$ at $q_x = \pm \pi$ or $q_y = \pm \pi$, and the corresponding equations for the wavepacket's motion under quadratic potentials are given by
\begin{align}
\dot{R}_x &= \frac{1}{2}E^{xx} \left\langle \frac{\partial g_{xx}}{\partial q^x} \right\rangle_{\mathcal{S}} + \frac{1}{2} E^{xy} \left\langle \frac{\partial g_{xy}}{\partial q^x} \right\rangle_{\mathcal{S}} \,, \label{eq:2D checkerboard dX} \\
\dot{R}_y &= \frac{1}{2}E^{xx} \left\langle \frac{\partial g_{xx}}{\partial q^y} \right\rangle_{\mathcal{S}} + \frac{1}{2}E^{xy} \left\langle \frac{\partial g_{xy}}{\partial q^y} \right\rangle_{\mathcal{S}} \,. \label{eq:2D checkerboard dY}
\end{align}
For a quadratic parabolic potential $V_{\mathrm{ext}} = E^{xx} x^2/2 $, the time evolution of the momentum space variance can be estimated approximately, as before, as 
\begin{equation}
\Delta W_{Q}^{xx} \approx (E^{xx})^2 W_{R,xx} t^2.
\end{equation}
Similarly, for a quadratic hyperbolic potential $V_{\mathrm{ext}} = E^{xy} xy/2 $, the theoretical distribution in momentum space can be estimated approximately as
\begin{equation}
\begin{split}
\Delta W_{Q}^{xx} &\approx (E^{xy})^2 W_{R,yy} t^2, \cr
\Delta W_{Q}^{yy} &\approx (E^{xy})^2 W_{R,xx} t^2, \cr
\end{split}
\end{equation}
From Eqs.~\eqref{eq:2D checkerboard dX}-\eqref{eq:2D checkerboard dY}, we observe that the velocity is only related to the quantum metric dipoles $\partial_\mu g_{\nu\rho}$.
In Fig.~\ref{fig:2D checkerboard}(a-d), we plot $\partial_\mu g_{\nu\rho}$ across the Brillouin zone that shows that this model allows us to investigate all the possible metric-induced displacements predicted by the semiclassical EOM.
In particular, notice that quadratic potentials of different type can give rise to distinct responses. 
A parabolic potential, $E^{xx}\neq 0$, yields a displacement analogous to a Hall effect through terms like $\dot R_y \sim E^{xx}\braket{\partial g_{xx}/\partial q^y}_{\mathcal{S}}$.
Instead, hyperbolic potentials, $E^{xy}\neq 0$, couple to both velocity directions.

To investigate these effects, we prepare wavepackets centered at different initial momenta. 
The first case of parabolic potential is analyzed for initial momentum $(Q^x, Q^y) = (\pi/a,\pi/3a)$, as indicated by a black cross in Fig.~\ref{fig:2D checkerboard}(a-b). 
This wavepacket shows a significant response along the $y$ direction when a quadratic potential $E^{xx}$ is applied, as shown in Fig.~\ref{fig:2D checkerboard}(e). Since $\partial g_{xx}/\partial q^{x}$ is anti-symmetric with respect to $q_x = \pi$, the response along the $x$ direction is zero thus producing an \emph{overall transverse response}.
For a hyperbolic potential, we prepare a wavepacket at $Q^x = Q^y = 2\pi/5a$, as shown by a black cross in Fig.~\ref{fig:2D checkerboard}(c-d). 
This wavepacket exhibits a response both in the $x$ and $y$ directions of opposite sign due to the different signs of the quantum metric dipole, as shown in Fig.~\ref{fig:2D checkerboard}(f). 
In the insets of Figs.~\ref{fig:2D checkerboard}(e-f), we also show the dispersion of the wavepacket in momentum space, displaying different behavior in the two cases analyzed here. 
The figure also shows the density distribution on different sublattices, and we observe that the displacement is actually originating from small deformations in each sublattice occupation. 
Since the deformation is not a real displacement, for a fully filled band under a quadratic potential, there is no contribution from the quantum metric to transport. 
This can also be understood from the fact that the contribution of the metric to the velocity takes the form of a total derivative and $\int_{\mathbb{T}^n} d^n\mathbf{k}~\partial_\mu g_{\nu\rho} = 0$.

\section{Discussion and summary}
\label{sec:4}

In this work we have demonstrated that dynamical effects can take place in flat bands despite the absence of dispersions. 
We have shown that the interplay of inhomogeneous potentials and band geometry are responsible for the appearance of Bloch oscillations in the mean position and variance of a wavepacket, as well as for the transverse displacement without a Berry curvature. 
We have carefully characterized these effects based on numerical simulations and a comparison with semiclassical equations of motion, which we have generalized to the non-Abelian case as well, finding excellent agreement.
It is worth stressing that we have derived the relevant equations using an operator approach that allows to obtain the final results with little algebraic effort, as explained in the Supplemental Material.

While we have investigated these effects in specific tight-binding models with flat bands, our results  are relevant to generic quantum systems displaying nearly flat bands, such as Moiré materials \cite{Levitov2021} or synthetic matter systems like photonic ones or cold atoms.
Notice that the displacements identified can require a high precision measurement of the wavepacket dynamics.
This could be an interesting testbed for recent advances in quantum gas microscopy with sub-lattice resolution \cite{Asteria2021}. 
A distinct approach to flat Bloch oscillations was explored in Ref.~\cite{Zeng2025} where these arise in the presence of strong fields via homogeneous electric fields and nonlinear response.
Interesting perspectives are to investigate effects in non-Abelian topological models based on the equations discussed in this work in the spirit of topological Bloch oscillations \cite{Alexandradinata, Marco-ncs41467-020-19518-x}, to study the regime on nonlinear response or generalize our results for inhomogenous time-dependent strain that behaves as an inhomogenous (pseudo)electric field \cite{Vozmediano2010}.

\section{Acknowledgment}

The authors acknowledge support from the  China Scholarship Council (CSC), the Quantum Technology Flagship project PASQuanS2, the INFN project Iniziativa Specifica IS-Quantum, the Italian Ministry of University and Research via the Rita Levi-Montalcini program.

\bibliographystyle{apsrev4-2}
\bibliography{REF}

\onecolumngrid
\newpage

\pagebreak
\widetext

\newcommand{\beginsupplement}{%
        \setcounter{equation}{0}
        \renewcommand{\theequation}{S\arabic{equation}}%
        \setcounter{figure}{0}
        \renewcommand{\thefigure}{S\arabic{figure}}%
        \setcounter{section}{0}
        \renewcommand{\thesection}{S\arabic{section}}
     }
\beginsupplement

\onecolumngrid

\begin{center} 
\textbf{
Geometric Bloch oscillations and transverse displacement \\  in flat band systems \\ -- Supplemental Material --
}
\end{center}

\maketitle

\section{Matrix elements in Bloch basis}

In this section, we introduce the mathematical framework used to construct the formalism employed to derive the equations presented in the main text. We also present the definition of the rank-3 tensor within our formalism. These constructions offer a compact and systematic approach for analyzing dynamics in both Abelian and non-Abelian case.

\subsection{Definition of projected expectation} 

We start considering a wavepacket state $\ket{\Psi} = \frac{1}{(2\pi)^D}\sum_n \int \mathrm{d}^D\mathbf{q}~c_n(\mathbf{q}) \ket{\psi^{(n)}_{\mathbf{q}}}$ where $\ket{\psi^{(n)}_{\mathbf{q}}}$ are orthonormal basis states labeled by a band index $n$ and quasi-momentum $\mathbf{q}$. 
Let the opertaor $\hat{O}$ be defined by
\begin{equation}
\hat{O} = \frac{1}{(2\pi)^D}\sum_{m,n} \int \mathrm{d}^D\mathbf{q}~\mathcal{O}_{mn}(\mathbf{q})\ket{\psi^{(m)}_{\mathbf{q}}} \bra{\psi^{(n)}_{\mathbf{q}}} \,,
\end{equation}
where $\mathcal{O}_{mn}$ are the matrix elements of $\hat{O}$ in the $\{\psi^{(n)}_{\mathbf{q}}\}$ basis. 
Notice that $ \int \mathrm{d}^D\mathbf{r}~e^{i (\mathbf{q}^\prime - \mathbf{q}) \cdot \mathbf{r}} = (2\pi)^D \delta(\mathbf{q} - \mathbf{q}^\prime)$.
The expectation value of $\hat{O}$ in the subspace $\mathcal{S}$ (a subset of bands) is defined by $\braket{\hat{O}}_{\mathcal{S}} \equiv \bra{\Psi} \hat{P}_{\mathcal{S}} \hat{O} \hat{P}_{\mathcal{S}} \ket{\Psi} $ where $\hat{P}_\mathcal{S}$ is the projection operator onto the subspace $\mathcal{S}$.\\
For operators involving commutator of covariant derivatives $\hat{D}_\mu = \partial_\mu - i \hat{A}_\mu$, with $\partial_\mu \equiv \partial/\partial q^\mu$, we have
\begin{equation}
\begin{split}
\left\langle \left[\hat{D}_\mu, \hat{O} \right] \right\rangle_{\mathcal{S}} &=\frac{1}{(2\pi)^D} \sum_{m,n\in \mathcal{S}} \int \mathrm{d}^D\mathbf{q}~c^\ast_m(\mathbf{q}) c_n(\mathbf{q}) \left( \frac{\partial}{\partial q^\mu} \mathcal{O}_{mn}(\mathbf{q})-i \left[ A_{\mu}(\mathbf{q}), O(\mathbf{q}) \right]_{mn} \right) \,.
\end{split}
\end{equation}
To simplify our calculations, we use the following standard identities involving the Dirac $\delta$-function and its derivatives
\begin{align}
\frac{\partial}{\partial q} \delta(q-q') &= - \frac{\partial}{\partial q^\prime} \delta(q-q^\prime) \,,\\
\int \mathrm{d}q~f(q)\frac{\partial}{\partial q} \delta(q-q^\prime) &= -\frac{\partial}{\partial q^\prime} f(q^\prime) \,. \label{eq: int delta}
\end{align}
The first identity reflects the antisymmetry of the derivative of the Dirac $\delta$-function under interchange of its arguments, while the second follows from integration by parts and the defining property of the Dirac $\delta$-function as a distribution.

\subsection{Matrix elements of the position operator in the Bloch basis}

The position operator, when acting on Bloch states, takes the form
\begin{equation}
\label{eq:rmu}
\hat{r}_\mu \ket{\psi_{\mathbf{q}}} = r_\mu e^{i \mathbf{q} \cdot \mathbf{r}} \ket{u_{\mathbf{q}}} = -i \partial_\mu \ket{\psi_{\mathbf{q}}} + i e^{i \mathbf{q} \cdot \mathbf{r}} \partial_\mu \ket{u_{\mathbf{q}}} \,.
\end{equation}
Based on this, we can get the following recursive equation for the operator $\hat{\mathcal{R}}^{(n)}_\mu \equiv \hat r_\mu^n e^{i\mathbf{q} \cdot \mathbf{r}}$
\begin{equation}
\begin{split}
\hat{\mathcal{R}}^{(n)}_\mu \ket{u_{\mathbf{q}}} & = (-i)^n (\partial^n_\mu e^{i\mathbf{q} \cdot \mathbf{r} })\ket{u_{\mathbf{q}}} \cr 
&= (-i\partial_\mu) ( r_\mu^{n-1}  e^{i\mathbf{q} \cdot \mathbf{r}} ) \ket{u_{\mathbf{q}}} - ( r_\mu^{n-1} e^{i\mathbf{q} \cdot \mathbf{r}} ) (-i \partial_\mu) \ket{u_{\mathbf{q}}} =  \cr
&= -i [\partial_\mu, \hat{\mathcal{R}}^{(n-1)}_\mu ] \ket{u_{\mathbf{q}}} \,,
\end{split}
\label{eq: Recursive relation}
\end{equation}
where $\hat{\mathcal{R}}^{(0)} = e^{i \mathbf{q} \cdot \mathbf{r} }$. Similarly as Eq.~\eqref{eq:rmu}, one can derive analogous equations for products of position operators in different directions (e.g. $\hat{r}_{\mu}\hat{r}_{\nu}$). Indeed, to calculate the semiclassical equations based on the projected position operator $\hat{P}_{\mathcal{S}} \left( \prod_j\hat{r}_j \right) \hat{P}_{\mathcal{S}}$, we need to express $\hat{r}_\mu$, as well as its products $\hat{r}_{\mu}\hat{r}_{\nu}$ and $\hat{r}_{\mu}\hat{r}_{\nu} \hat{r}_\rho$, in the Bloch basis. 
From Eq.~\eqref{eq:rmu} and its generalization to higher powers (not indicated here), we obtain
\begin{align}
\bra{\psi^{(m)}_{\mathbf{q}}} \hat{r}_\mu \ket{\psi^{(n)}_{\mathbf{q}^\prime}} &= -i \delta_{mn} \frac{\partial}{\partial q_\mu^\prime} \delta(\mathbf{q} - \mathbf{q}^\prime) + \frac{i}{(2\pi)^D} \int \mathrm{d}^D\mathbf{r}~e^{i (\mathbf{q}^\prime - \mathbf{q}) \cdot \mathbf{r}} \left\langle u^{(n)}_{\mathbf{q}} \right| \frac{\partial}{\partial q_\mu^\prime} \left| u^{(n)}_{\mathbf{q}^\prime} \right\rangle \nonumber \\
&= \delta(\mathbf{q} - \mathbf{q}^\prime)\left[ i \delta_{mn} \frac{\partial}{\partial q_\mu^\prime} + A^{(m,n)}_\mu\right] 
\label{eq:r1 matrix element} \,, \\
\bra{\psi^{(m)}_\mathbf{q}} \hat{r}_\mu \hat{r}_\nu \ket{\psi^{(n)}_{\mathbf{q}^\prime}} &= \delta(\mathbf{q} - \mathbf{q}^\prime)\left[- \delta_{mn} \frac{\partial}{\partial q_\mu^\prime} \frac{\partial}{\partial q_\nu^\prime} + i A^{(m,n)}_\nu  \frac{\partial}{\partial q_\mu^\prime} + i  A^{(m,n)}_\mu  \frac{\partial}{\partial q_\nu^\prime} - \left\langle u^{(m)}_\mathbf{q} \right. \left| \partial_\mu \partial_\nu u^{(n)}_{\mathbf{q}} \right\rangle \right] \,, 
\label{eq:r2 matrix element} \\
\bra{\psi^{(n)}_\mathbf{q}} \hat{r}_\mu \hat{r}_\nu \hat{r}_\rho \ket{\psi^{(n)}_{\mathbf{q}^\prime}} &= -i\delta(\mathbf{q} - \mathbf{q}^\prime) \left[ \delta_{mn}\frac{\partial}{\partial q_\mu^\prime} \frac{\partial}{\partial q_\nu^\prime}\frac{\partial}{\partial q_\rho^\prime} - i A^{(m,n)}_\mu \frac{\partial}{\partial q_\nu^\prime} \frac{\partial}{\partial q_\rho^\prime} - i A_{\nu}^{(m,n)} \frac{\partial}{\partial q_\rho^\prime} \frac{\partial}{\partial q_\mu^\prime} - i A_\rho^{(m,n)} \frac{\partial}{\partial q_\mu^\prime} \frac{\partial}{\partial q_\nu^\prime} \right. \nonumber \\
&\left. + \left\langle u^{(m)}_{\mathbf{q}} \right. \left| \partial_\mu \partial_\nu u^{(n)}_{\mathbf{q}} \right\rangle \frac{\partial}{\partial q_\rho^\prime} + \left\langle u^{(m)}_{\mathbf{q}} \right. \left| \partial_\nu \partial_\rho u^{(n)}_{\mathbf{q}} \right\rangle \frac{\partial}{\partial q_\mu^\prime} + \left\langle u^{(m)}_{\mathbf{q}} \right. \left| \partial_\rho \partial_\mu u^{(n)}_{\mathbf{q}} \right\rangle \frac{\partial}{\partial q_\nu^\prime} + \left\langle u^{(m)}_{\mathbf{q}} \right. \left| \partial_\mu \partial_\nu \partial_\rho u^{(n)}_{\mathbf{q}} \right\rangle \right] \,,
\label{eq:r3 matrix element}
\end{align}
where we employed the relation in Eq.~\eqref{eq: int delta}.
In this form, these matrix elements can be interpreted as representing an operator that acts uniquely on the wavefunction on the right-hand side.
The matrix element $\bra{u_{\mathbf{q}}^{(m)}} \partial_\mu \partial_\nu \ket{u_{\mathbf{q}}^{(n)}}$ can be expressed in terms of the Berry connection and the quantum metric tensor as follows
\begin{equation}
\begin{split}
\braket{u^{(m)}_\mathbf{q}| \frac{\partial}{\partial q_\mu} \frac{\partial}{\partial q_\nu} u^{(n)}_{\mathbf{q}}} = -\frac{i}{2} \left(\partial_\mu A_\nu^{(m,n)} + \partial_\nu A_\mu^{(m,n)} \right) - g^{(m,n)}_{\mu\nu} - \frac{1}{2} \sum_l\left(A^{(m,l)}_{\mu} A^{(l,n)}_{\nu} + A^{(m,l)}_{\nu} A^{(l,n)}_{\mu} \right) \,. \cr
\end{split}
\end{equation}
To simplify the calculations, we introduce the following notation to represent the projected operators in the subspace $\mathcal{S}$: $\hat{r}_{\mu,\mathcal{S}} \equiv \hat{P}_{\mathcal{S}} \hat{r}_\mu \hat{P}_{\mathcal{S}}$, $\hat{r}^{(2)}_{\mu\nu,\mathcal{S}} \equiv \hat{P}_{\mathcal{S}} \hat{r}_\mu \hat{r}_\nu \hat{P}_{\mathcal{S}}$ and $\hat{r}^{(3)}_{\mu\nu\rho,\mathcal{S}} \equiv \hat{P}_{\mathcal{S}} \hat{r}_\mu \hat{r}_\nu \hat{r}_\rho \hat{P}_{\mathcal{S}}$. 
These operators are defined via Eqs.~\eqref{eq:r1 matrix element}, \eqref{eq:r2 matrix element} and \eqref{eq:r3 matrix element}, which correspond to the multipole moment operator and its products in the Bloch basis. 
This approach allows us to obtain the following expressions
\begin{align}
\hat{r}_{\mu,\mathcal{S}} &:= i\partial_\mu + \hat{\mathbf{A}}_\mu = i \hat{\boldsymbol{D}}_\mu \,, \label{eq: r1 operator} \\
\hat{r}^{(2)}_{\mu\nu,\mathcal{S}} &:= \frac{1}{2} \left\{ \hat{r}_{\mu,\mathcal{S}}, \hat{r}_{\nu,\mathcal{S}} \right\} + \hat{\boldsymbol{g}}_{\mu\nu} \,,  \label{eq: r2 operator} \\
\hat{r}^{(3)}_{\mu\nu\rho,\mathcal{S}} &:= \frac{1}{12} \{\hat{r}_{\mu,\mathcal{S}},\{\hat{r}_{\nu,\mathcal{S}}, \hat{r}_{\rho,\mathcal{S}} \} \} + \frac{1}{12}\{\hat{r}_{\nu,\mathcal{S}},\{\hat{r}_{\rho,\mathcal{S}}, \hat{r}_{\mu,\mathcal{S}} \} \} + \frac{1}{12}\{\hat{r}_{\rho,\mathcal{S}}, \{\hat{r}_{\mu,\mathcal{S}}, \hat{r}_{\nu,\mathcal{S}} \} \} + \frac{1}{2} \{ \hat{\boldsymbol{g}}_{\mu\nu}, \hat{r}_{\rho,\mathcal{S}} \} \cr
&+ \frac{1}{2} \{ \hat{\boldsymbol{g}}_{\nu\rho}, \hat{r}_{\mu,\mathcal{S}}\} + \frac{1}{2} \{ \hat{\boldsymbol{g}}_{\rho\mu}, \hat{r}_{\nu,\mathcal{S}} \} + \hat{\boldsymbol{T}}_{\mu\nu\rho} \,,
\label{eq: r3 operator} 
\end{align}
$\boldsymbol{D}_{\mu}$ is the covariant derivative, while the Berry connection, Berry curvature and quantum metric are respectively given by
\begin{align}
\hat{\mathbf{A}}_\mu(\mathbf{q}) &= i\sum_{m,n\in\mathcal{S}} \ket{\psi^{(m)}_{\mathbf{q}}} \braket{u^{(m)}_{\mathbf{q}}|\partial_\mu u^{(n)}_{\mathbf{q}}} \bra{\psi^{(n)}_{\mathbf{q}}} = \sum_{m,n \in\mathcal{S}} A^{(m,n)}_{\mu}(\mathbf{q})\ket{\psi^{(m)}_{\mathbf{q}}} \bra{\psi^{(n)}_{\mathbf{q}}} \,,\\
\hat{\boldsymbol{\Omega}}_{\mu\nu}(\mathbf{q}) &= i\sum_{m,n\in\mathcal{S}} \ket{\psi^{(m)}_{\mathbf{q}}} \left[ \braket{ \partial_\mu u^{(m)}_{\mathbf{q}}|\partial_\nu u^{(n)}_{\mathbf{q}}} 
 - \braket{ \partial_\nu u^{(m)}_{\mathbf{q}}|\partial_\mu u^{(n)}_{\mathbf{q}}} -i \sum_{l \in \mathcal{S}} \left( A_{\mu}^{(m,l)} A^{(l,n)}_{\nu} - i A^{(m,l)}_\nu A^{(l,n)}_{\mu} \right) \right] \bra{\psi^{(n)}_{\mathbf{q}}} \nonumber\\
&= \sum_{m,n \in \mathcal{S}} \Omega^{(m,n)}_{\mu\nu}(\mathbf{q}) \ket{\psi^{(m)}_{\mathbf{q}}} \bra{\psi^{(n)}_{\mathbf{q}}} \,, \\
\hat{\boldsymbol{g}}_{\mu\nu}(\mathbf{q}) &= \sum_{m,n \in \mathcal{S}} \ket{\psi^{(m)}_{\mathbf{q}}} \left[ \frac{1}{2}\braket{ \partial_\mu u^{(m)}_{\mathbf{q}}|\partial_\nu u^{(n)}_{\mathbf{q}}} 
 - \frac{1}{2} \sum_{l\in \mathcal{S}} \braket{ \partial_\nu u^{(m)}_{\mathbf{q}}| u^{(l)}_{\mathbf{q}}}  \braket{ u^{(l)}_{\mathbf{q}}|\partial_\mu u^{(n)}_{\mathbf{q}}} + (\mu \leftrightarrow \nu) \right] \bra{\psi^{(n)}_{\mathbf{q}}} \nonumber \\
 &= \sum_{m,n \in \mathcal{S}} g^{(m,n)}_{\mu\nu}(\mathbf{q}) \ket{\psi^{(m)}_{\mathbf{q}}} \bra{\psi^{(n)}_{\mathbf{q}}} \,.
\end{align}

\subsection{Gauge invariant rank-3 tensor} 

The rank-3 tensor $\hat{\boldsymbol{T}}_{\mu\nu\rho}$~\cite{Kozii-PhysRevLett.126.156602,Ahn-PhysRevX.10.041041,Ahn-2022NatPhys,Xu2025} is defined by decomposing the matrix elements of $\hat{r}^{(3)}_{\mu\nu\rho,\mathcal{S}}$ and subtracting the components that can be independently constructed from $\hat{r}_{\mu,\mathcal{S}}$ and $ \hat{r}^{(2)}_{\mu\nu,\mathcal{S}}$.
In our calculation, the tensor $\boldsymbol{T}_{\mu\nu\rho}$ can be decomposed into Abelian and non-Abelian components as $\boldsymbol{T}_{\mu\nu\rho} = (\boldsymbol{T}_{\mu\nu\rho})_{\mathrm{Abelian}} + (\boldsymbol{T}_{\mu\nu\rho})_\mathrm{non-Abelian}$ with contributions expressed in terms of the matrix elements $\braket{u^{(m)}|\partial_\mu\partial_\nu \partial_\rho u^{(n)}}$, $ \boldsymbol{g}_{\mu\nu}$ and $\boldsymbol{D}_{\mu}$
\begin{align}
(\hat{\boldsymbol{T}}_{\mu\nu\rho})_{\mathrm{Abelian}} &= -\frac{i}{4}\left( \left[ \hat{\boldsymbol{D}}_\mu, 2\hat{\boldsymbol{g}}_{\nu\rho} + \left\{ \hat{\mathbf{A}}_\nu, \hat{\mathbf{A}}_\rho \right\} \right] + \left[ \hat{\boldsymbol{D}}_\nu, 2 \hat{\boldsymbol{g}}_{\rho\mu} + \left\{ \hat{\mathbf{A}}_\rho, \hat{\mathbf{A}}_\mu \right\} \right] + \left[ \hat{\boldsymbol{D}}_\rho, 2 \hat{\boldsymbol{g}}_{\mu\nu} + \left\{ \hat{\mathbf{A}}_\mu, \hat{\mathbf{A}}_\nu \right\} \right] \right) \nonumber \\
& + \frac{1}{3} \left( \left[ \hat{\boldsymbol{D}}_\mu, \left[ \hat{\boldsymbol{D}}_\rho, \hat{\mathbf{A}}_\nu \right] \right] + \left[ \hat{\boldsymbol{D}}_\nu, \left[ \hat{\boldsymbol{D}}_\mu, \hat{\mathbf{A}}_\rho \right] \right] + \left[ \hat{\boldsymbol{D}}_\rho, \left[ \hat{\boldsymbol{D}}_\nu, \hat{\mathbf{A}}_\mu \right] \right] \right) \nonumber \\
& - \frac{1}{12} \left( \left\{ \left\{ \hat{\mathbf{A}}_\mu, \hat{\mathbf{A}}_\nu \right\}, \hat{\mathbf{A}}_\rho \right\} + \left\{ \left\{ \hat{\mathbf{A}}_\nu, \hat{\mathbf{A}}_\rho \right\}, \hat{\mathbf{A}}_\mu \right\} + \left\{ \left\{ \hat{\mathbf{A}}_\rho, \hat{\mathbf{A}}_\mu \right\}, \hat{\mathbf{A}}_\nu \right\} \right) \nonumber \\
&- \frac{1}{2} \left( \left\{ \hat{\boldsymbol{g}}_{\mu\nu} ,\hat{\mathbf{A}}_\rho \right\} + \left\{\hat{\boldsymbol{g}}_{\nu\rho}, \hat{\mathbf{A}}_\mu \right\} + \left\{ \hat{\boldsymbol{g}}_{\rho\mu}, \hat{\mathbf{A}}_\nu \right\} \right) -i \sum_{m,n\in \mathcal{S}}\braket{u^{(m)}|\partial_\mu \partial_\nu \partial_\rho u^{(n)}} \,, \label{eq: rank-3 Abelian} \\
(\hat{\boldsymbol{T}}_{\mu\nu\rho})_{\mathrm{non-Abelian}} &= \frac{i}{3} \left[ \hat{\mathbf{A}}_\rho,\left[ \hat{\boldsymbol{D}}_\mu, \hat{\mathbf{A}}_\nu \right] \right] + \frac{i}{3} \left[ \hat{\mathbf{A}}_\mu, \left[\hat{\boldsymbol{D}}_\nu, \hat{\mathbf{A}}_\rho \right] \right] + \frac{i}{3} \left[ \hat{\mathbf{A}}_\nu, \left[ \hat{\boldsymbol{D}}_\rho, \hat{\mathbf{A}}_\mu \right] \right] \nonumber \\
&+ \frac{1}{6} \left[ \hat{\mathbf{A}}_\mu, \left\{ \hat{\mathbf{A}}_\nu, \hat{\mathbf{A}}_\rho \right\} \right] + \frac{1}{6} \left[ \hat{\mathbf{A}}_\nu, \left\{ \hat{\mathbf{A}}_\rho, \hat{\mathbf{A}}_\mu \right\} \right] + \frac{1}{6} \left[ \hat{\mathbf{A}}_\rho, \left\{\hat{\mathbf{A}}_\mu, \hat{\mathbf{A}}_\nu \right\} \right] \nonumber \\
& - \frac{i}{12} \left[ \hat{\mathbf{A}}_\mu, \left[ \hat{\boldsymbol{D}}_\nu, \hat{\mathbf{A}}_\rho \right] + \left[ \hat{\boldsymbol{D}}_\rho, \hat{\mathbf{A}}_\nu \right] \right] - \frac{i}{12} \left[ \hat{\mathbf{A}}_\nu, \left[ \hat{\boldsymbol{D}}_\rho, \hat{\mathbf{A}}_\mu \right] + \left[ \hat{\boldsymbol{D}}_\mu, \hat{\mathbf{A}}_\rho \right] \right] \nonumber \\
&- \frac{i}{12} \left[ \hat{\mathbf{A}}_\rho, \left[ \hat{\boldsymbol{D}}_\mu, \hat{\mathbf{A}}_\nu \right] + \left[ \hat{\boldsymbol{D}}_\nu, \hat{\mathbf{A}}_\mu \right] \right] - \frac{1}{2} \left[ \hat{\boldsymbol{g}}_{\mu\nu} ,\hat{\mathbf{A}}_\rho \right] - \frac{1}{2} \left[ \hat{\boldsymbol{g}}_{\nu\rho}, \hat{\mathbf{A}}_\mu \right] - \frac{1}{2} \left[ \hat{\boldsymbol{g}}_{\rho\mu}, \hat{\mathbf{A}}_\nu \right] \,. \label{eq: rank-3 non-Abelian}
\end{align}
Notice that Eq.~\eqref{eq: rank-3 Abelian} simplifies in the Abelian case presented in Ref.~\cite{Kozii-PhysRevLett.126.156602}. 
The non-Abelian contribution, given in Eq.~\eqref{eq: rank-3 non-Abelian}, vanishes for a single band as it is made only of commutators and therefore justifies our naming of it as non-Abelian.
Moreover, there exists an important relationship between the tensor $\hat{\boldsymbol{T}}_{\mu\nu\rho}$ and the quantum geometric connection, which is given by \cite{Ahn-PhysRevX.10.041041,Kozii-PhysRevLett.126.156602,Avdoshkin-PhysRevB.107.245136,mitscherling2024gaugeinvariantprojectorcalculusquantum}
\begin{equation}
c_{\mu\nu\rho} = \mathrm{Tr}[ \hat{P}_\mathcal{S} (\partial_\mu \partial_\nu \hat{P}_\mathcal{S}) (\partial_\rho \hat{P}_\mathcal{S})] \,,
\label{eq: geometric connection}
\end{equation}
where $\hat{P}_\mathcal{S}$ denotes the projection operator onto the occupied subspace. 
This geometric connection tensor $c_{\mu\nu\rho}$ is symmetric under the exchange $\mu \leftrightarrow\nu$, indicating that it defines a torsion-free connection. 
To show the relation with the rank-3 tensor, it is convenient to treat the Abelian case and split $c_{\mu\nu\rho}$ as the sum of two contributions
\begin{align}
c^{(1)}_{\mu\nu\rho} &= -i (A_{\mu} g_{\nu\rho} + A_{\nu} g_{\mu\rho} + A_{\rho} g_{\mu\nu} ) + \frac{3}{4}\partial_\nu (g_{\mu\rho} + A_{\mu} A_{\rho}) + \frac{3}{4}\partial_\mu ( g_{\nu\rho} + A_\nu A_\rho) -i A_{\mu} A_{\nu} A_{\rho} \nonumber\\
&+ i \partial_\mu \partial_\nu A_\rho + \braket{ u| \partial_\mu \partial_\nu \partial_\rho u} \,, \\
c^{(2)}_{\mu\nu\rho} &= - \frac{i}{2} (\partial_\nu \Omega_{\mu\rho} + \partial_\mu \Omega_{\nu \rho}) -A_\rho \partial_\mu A_\nu - \frac{1}{2}( A_{\mu} \Omega_{\nu\rho} - A_{\nu} \Omega_{\rho\mu} - A_\rho \Omega_{\mu\nu}) + \frac{1}{4}\partial_\nu (g_{\rho\mu} + A_{\rho} A_{\mu}) \nonumber\\
&+ \frac{1}{4}\partial_\mu ( g_{\nu\rho} + A_\nu A_\rho) \,.
\end{align}
The cyclic permutation sum of $c^{(1)}_{\mu\nu\rho}$ is related to the gauge invariant rank-3 tensor $\boldsymbol{T}_{\mu\nu\rho}$ through the following relation
\begin{equation}
\begin{split}
i \boldsymbol{T}_{\mu\nu\rho} &= \frac{1}{3} \left( c^{(1)}_{\mu\nu\rho} + c^{(1)}_{\nu\rho\mu} + c^{(1)}_{\rho\mu\nu} \right) \cr
&= -i( A_\mu g_{\nu\rho} + A_{\nu} g_{\mu\rho} + A_{\rho} g_{\mu\nu} ) + \frac{1}{2} \partial_\mu \left( g_{\nu\rho} + A_{\nu} A_\rho \right) + \frac{1}{2}\partial_\nu \left( g_{\rho\mu} + A_{\rho} A_\mu \right) +  \frac{1}{2}\partial_\rho \left( g_{\mu\nu} + A_{\mu} A_\nu \right)  \cr
& + \frac{i}{3} (\partial_\mu \partial_\nu A_\rho + \partial_\nu \partial_\rho A_\mu + \partial_\rho \partial_\mu A_\nu) -i A_\mu A_\nu A_\rho + \braket{u| \partial_\mu \partial_\nu \partial_\rho u}\,,    
\end{split}
\end{equation}
and $\boldsymbol{T}_{\mu\nu\rho}$ as defined above is purely real. 
The cyclic permutation sum of the second part, $c^{(2)}_{\mu\nu\rho}$, yields another real and gauge-invariant quantity
\begin{equation}
\frac{1}{3} \left( c^{(2)}_{\mu\nu\rho} + c^{(2)}_{\nu\rho\mu} + c^{(2)}_{\rho\mu\nu} \right) = \frac{1}{6} \left( \partial_\mu g_{\nu\rho} + \partial_\nu g_{\rho\mu} + \partial_\rho g_{\mu\nu} \right) \,.  
\end{equation}
such that
\begin{equation}
\begin{split}
c_{\mu\nu\rho} - i \boldsymbol{T}_{\mu\nu\rho} 
&= - \Gamma_{\mu\nu\rho} + \frac{i}{2} \tilde{\Gamma}_{\mu\nu\rho} \,,
\end{split}
\end{equation}
where $\Gamma_{\mu\nu\rho}$ is identified with the Christoffel symbol
\begin{equation}
\Gamma_{\mu\nu\rho} = g_{\rho\lambda}\Gamma_{\mu\nu}^\lambda = \frac{1}{2} \left( \partial_\mu g_{\nu\rho} + \partial_\nu g_{\rho\mu} - \partial_\rho g_{\mu\nu} \right) \,,
\end{equation}
while $\tilde{\Gamma}_{\mu\nu\rho}$ is a symplectic connection
\begin{equation}
\tilde{\Gamma}_{\mu\nu\rho} = \Omega_{\rho\lambda}\tilde{\Gamma}_{\mu\nu}^{\lambda} = \frac{1}{3} \left( \partial_\mu \Omega_{\rho\nu} + \partial_\nu \Omega_{\rho\mu} \right) \,.
\end{equation}
Our results are consistent with Ref.~\cite{Kozii-PhysRevLett.126.156602,Xu2025} and for higher-order matrix elements $\hat{r}^{(n)}_{\mu\nu\cdots,~\mathcal{S}}$, the same method can be systematically applied to define and compute the corresponding rank-$n$ gauge invariant tensors.

\section{Building semiclassical equations via projected Heisenberg equation}

In this section, we provide the detailed derivation of the full semiclassical dynamics of the wavepacket’s center and variance using the formalism developed above.
We consider a Hamiltonian of the form
\begin{equation}
\hat{H} = \hat H_0 + E^\mu \hat{r}_\mu + \frac{1}{2} E^{\mu\nu} \hat{r}_\mu \hat{r}_\nu + \frac{1}{6} E^{\mu\nu\rho} \hat{r}_\mu \hat{r}_\nu \hat{r}_\rho\,,
\end{equation}
and derive the projected Heisenberg equations $\partial_t\braket{{\hat O}}_{\mathcal{S}} = -i \braket{[\hat O, \hat H]}_{\mathcal{S}}$ for each observable.

\subsection{Semiclassical equations of wavepacket velocity}

Here we derive the semiclassical equations for the velocity, and thus we take $\hat O \equiv \hat r_\mu$ for the Heisenberg equations. 
Note that we need to compute commutators with all the terms appearing in the inhomogeneous potential.
Therefore, by employing Eqs.~\eqref{eq: r1 operator},~\eqref{eq: r2 operator} and ~\eqref{eq: r3 operator}, the commutators appearing in the projected Heisenberg equation can be calculated straightforwardly
\begin{align}
\left[ \hat{r}_{\mu,\mathcal{S}}, \hat{r}_{\nu,\mathcal{S}} \right] &= \left[ i\hat{\boldsymbol{D}}_\mu, i\hat{\boldsymbol{D}}_\nu \right] = i\hat{\boldsymbol{\Omega}}_{\mu\nu} 
\label{eq:[r,r]} \,, \\
\left[ \hat{r}_{\mu,\mathcal{S}}, \hat{r}^{(2)}_{\nu\rho,\mathcal{S}} \right] &= \frac{i}{2} \left\{ \hat{r}_{\nu,\mathcal{S}}, \hat{\boldsymbol{\Omega}}_{\mu\rho} \right\} + \frac{i}{2} \left\{ \hat{r}_{\rho,\mathcal{S}}, \hat{\boldsymbol{\Omega}}_{\mu\nu} \right\} + i \left[ \hat{\boldsymbol{D}}_\mu, \hat{\boldsymbol{g}}_{\nu\rho} \right] \,,
\label{eq:[r,rr]} \\
\left[ \hat{r}_{\mu,\mathcal{S}}, \hat{r}^{(3)}_{\nu\rho\lambda,\mathcal{S}} \right] &= \frac{i}{4} \left\{ \left\{ \hat{r}_{\rho,\mathcal{S}}, \hat{r}_{\lambda,\mathcal{S}} \right\}, \hat{\boldsymbol{\Omega}}_{\mu\nu} \right\} + \frac{i}{4} \left\{ \left\{ \hat{r}_{\lambda,\mathcal{S}}, \hat{r}_{\nu,\mathcal{S}} \right\}, \hat{\boldsymbol{\Omega}}_{\mu\rho} \right\} + \frac{i}{4} \left\{ \left\{ \hat{r}_{\nu,\mathcal{S}}, \hat{r}_{\rho,\mathcal{S}} \right\}, \hat{\boldsymbol{\Omega}}_{\mu\lambda} \right\} + \left[i \hat{\boldsymbol{D}}_\mu, \boldsymbol{T}_{\nu\rho\lambda} \right] \nonumber \\
&+ \frac{i}{2} \left\{ \hat{\boldsymbol{g}}_{\nu\rho}, \hat{\boldsymbol{\Omega}}_{\mu\lambda} \right\} + \frac{i}{2} \left\{ \hat{\boldsymbol{g}}_{\rho\lambda}, \hat{\boldsymbol{\Omega}}_{\mu\nu} \right\} + \frac{i}{2} \left\{ \hat{\boldsymbol{g}}_{\lambda\nu}, \hat{\boldsymbol{\Omega}}_{\mu\rho} \right\} \nonumber \\
&+ \frac{i}{2} \left\{ \hat{r}_{\lambda,\mathcal{S}}, \left[ \hat{\boldsymbol{D}}_\mu, \hat{\boldsymbol{g}}_{\nu\rho} \right] \right\} + \frac{i}{2} \left\{ \hat{r}_{\nu,\mathcal{S}}, \left[ \hat{\boldsymbol{D}}_\mu, \hat{\boldsymbol{g}}_{\rho\lambda} \right] \right\}  + \frac{i}{2} \left\{ \hat{r}_{\rho,\mathcal{S}}, \left[ \hat{\boldsymbol{D}}_\mu, \hat{\boldsymbol{g}}_{\lambda\nu} \right] \right\} \,.
\label{eq:[r,rrr]}
\end{align}
To derive the above expressions, we have used the Jacobi identity (Eqs.~\eqref{eq: Jacobi identity 1} and \eqref{eq: Jacobi identity 2}), along with the relation between the commutators and anti-commutators (Eq. \eqref{eq: BAC})
\begin{align}
[A,[B,C]] + [B,[C,A]] + [C,[A,B]] & = 0 \,,
\label{eq: Jacobi identity 1}\\
[A,\{B,C\}] + [B,\{C,A\}] + [C,\{A,B\}] & = 0 \,,
\label{eq: Jacobi identity 2} \\
\{A,[B,C]\} + \{C,[B,A]\} &= [B,\{A,C\}] \,.
\label{eq: BAC}
\end{align}

Using the results above, we can derive the wavepacket velocity $\dot R_\mu \equiv \partial_t\braket{{\hat r_\mu}}_{\mathcal{S}}$ from the Heisenberg equations. We split the contributions from each potential term as
\begin{align}
\dot{R}_{\mu,0} &= \left\langle \left[ \hat{\boldsymbol{D}}_\mu,\hat H_0 \right] \right\rangle_{\mathcal{S}} \,, \label{eq:r0} \\
\dot{R}_{\mu,1} &= E^\nu \braket{\hat{\boldsymbol{\Omega}}_{\mu\nu}}_{\mathcal{S}} \,, \label{eq:r1} \\
\dot{R}_{\mu,2} &= \frac{1}{2} E^{\nu\rho} \left[ \frac{1}{2} \left\langle \left\{\hat{r}_{\nu,\mathcal{S}},\hat{\boldsymbol{\Omega}}_{\mu\rho}\right\} \right\rangle_{\mathcal{S}} + (\nu \leftrightarrow \rho) \right] + \frac{1}{2} E^{\nu\rho} \left\langle \left[ \hat{\boldsymbol{D}}_\mu, \hat{\boldsymbol{g}}_{\nu\rho} \right] \right\rangle_{\mathcal{S}} \,, \label{eq:r2} \\
\dot{R}_{\mu,3} &= \frac{1}{6} E^{\nu\rho\lambda} \left( \frac{1}{4} \left\langle \left\{ \hat{\boldsymbol{\Omega}}_{\mu\nu}, \left\{ \hat{r}_{\rho,\mathcal{S}}, \hat{r}_{\lambda,\mathcal{S}} \right\} \right\} \right\rangle_{\mathcal{S}} + \frac{1}{2} \left\langle \left\{ \left[ \hat{\boldsymbol{D}}_\mu, \hat{\boldsymbol{g}}_{\nu\rho} \right], \hat{r}_{\lambda,\mathcal{S}} \right\} \right\rangle_{\mathcal{S}} + \frac{1}{2} \left\langle \left\{ \hat{\boldsymbol{\Omega}}_{\mu\nu}, \hat{\boldsymbol{g}}_{\rho\lambda} \right\} \right\rangle_{\mathcal{S}}  + (\nu,\rho,\lambda)_P \right) \nonumber \\
& + \frac{1}{6} E^{\nu\rho\lambda} \left\langle \left[ \hat{\boldsymbol{D}}_\mu, \hat{\boldsymbol{T}}_{\nu\rho\lambda} \right] \right\rangle_{\mathcal{S}} \,. \label{eq:r3}
\end{align}
For inhomogeneous potentials of order higher than cubic, terms of the form $\{\{\{\hat{r}_{\mu,\mathcal{S}}, \hat{r}_{\nu,\mathcal{S}} \}, \hat{r}_{\rho,\mathcal{S}}\},\cdots\}$ will systematically appear.
In the case of a subspace $\mathcal S$ consisting of perfectly degenerate bands which are gapped from the others, the Hamiltonian $\hat H_0$ acts as the identity and the commutator $\braket{[\hat{O}, \hat H_0]}_{\mathcal{S}}$ vanishes, except when $\hat{O}$ is a derivative operator. 
Here, $(\nu,\rho,\lambda)_P$ represents the cyclic permutation of the indices $\nu,\rho,\lambda$.
In the main text, we have decomposed these equations into a shape-independent part and a shape-dependent part. The former reads
\begin{equation}
\begin{split}
\dot{R}_{\mu,\text{SI}} &= \left\langle [\hat{\boldsymbol{D}}_\mu, \hat{H}_0] \right\rangle_{\mathcal{S}} + E^\nu(\mathbf{R}) \left\langle \hat{\boldsymbol{\Omega}}_{\mu\nu} \right\rangle_{\mathcal{S}} + \frac{1}{2}E^{\nu\rho}(\mathbf{R}) \left\langle \left[ \hat{\boldsymbol{D}}_\mu, \hat{\boldsymbol{g}}_{\nu\rho} \right] \right\rangle_{\mathcal{S}} \cr
&+ \frac{1}{6} E^{\nu\rho\lambda} \left[ \frac{1}{2} \left\langle \left\{ \hat{\boldsymbol{\Omega}}_{\mu\nu}, \hat{\boldsymbol{g}}_{\rho\lambda} \right\} \right\rangle_{\mathcal{S}}  + (\nu,\rho,\lambda)_P \right] + \frac{1}{6} E^{\nu\rho\lambda} \left\langle \left[ \hat{\boldsymbol{D}}_\mu, \hat{\boldsymbol{T}}_{\nu\rho\lambda} \right] \right\rangle_{\mathcal{S}} \,, \cr
\end{split}
\end{equation}
where
\begin{align}
E^\nu(\mathbf{R}) &= E^\nu + \frac{1}{2} \left( E^{\nu\nu} \braket{\hat{r}_\nu}_{\mathcal{S}} + E^{\nu\rho} \braket{\hat{r}_\rho}_{\mathcal{S}} \right) + \frac{1}{6} \left( E^{\nu\rho\lambda} \braket{\hat{r}_\rho}_{\mathcal{S}} \braket{\hat{r}_\lambda}_{\mathcal{S}} + 2 E^{\nu\nu\rho} \braket{\hat{r}_\nu}_{\mathcal{S}} \braket{\hat{r}_\rho}_{\mathcal{S}} \right) \,, \label{eq:E1(R)} \\
E^{\nu\rho}(\mathbf{R}) &= E^{\nu\rho} + \frac{1}{6} \left( E^{\nu\rho\rho} \braket{\hat{r}_\rho}_{\mathcal{S}} +  E^{\nu\rho\lambda} \braket{\hat{r}_\lambda}_{\mathcal{S}} +  E^{\nu\nu\rho} \braket{\hat{r}_\nu}_{\mathcal{S}} \right) \label{eq:E2(R)} \,. 
\end{align}
Here, the effective field strength $E^\nu(\mathbf{R})$ and $E^{\nu\rho}(\mathbf{R})$ characterize the external field experienced by the wavepacket at position $\mathbf{R}$. It depends only on $\mathbf{R}=\braket{\hat{\mathbf{r}}}_{\mathcal{S}}$, the mean position of the wavepacket, and is therefore independent of its shape.

The shape-dependent part is instead given by
\begin{equation}
\label{eq:RSD}
\begin{split}
\dot{R}_{\mu,\text{SD}} &= \frac{1}{2} E^{\nu\rho} \left[ \frac{1}{2}\left\langle \left\{ \hat{r}_{\nu,\mathcal{S}}, \hat{\boldsymbol{\Omega}}_{\mu\rho} \right\} \right\rangle_{\mathcal{S}} - \braket{\hat{r}_\nu}_{\mathcal{S}} \left\langle \hat{\boldsymbol{\Omega}}_{\mu\rho} \right\rangle_{\mathcal{S}} + (\nu \leftrightarrow \rho) \right]\cr
& + \frac{1}{6} E^{\nu\rho\lambda} \left[ \frac{1}{4} \left\langle \left\{ \hat{r}_{\rho,\mathcal{S}}, \left\{ \hat{r}_{\lambda,\mathcal{S}} , \hat{\boldsymbol{\Omega}}_{\mu\nu} \right\} \right\} \right\rangle_{\mathcal{S}} - \braket{\hat{r}_\rho}_{\mathcal{S}} \braket{\hat{r}_\lambda}_{\mathcal{S}} \left\langle \hat{\boldsymbol{\Omega}}_{\mu\nu} \right\rangle_{\mathcal{S}}  + (\nu,\rho,\lambda)_P\right] \cr
& + \frac{1}{6} E^{\nu\rho\lambda} \left[ \frac{1}{2} \left\langle \left\{ \hat{r}_{\lambda,\mathcal{S}} , \left[ \hat{\boldsymbol{D}}_{\mu}, \hat{\boldsymbol{g}}_{\nu\rho} \right] \right\} \right\rangle_{\mathcal{S}} - \braket{\hat{r}_\lambda}_{\mathcal{S}} \left\langle \left[ \hat{\boldsymbol{D}}_{\mu}, \hat{\boldsymbol{g}}_{\nu\rho} \right] \right\rangle_{\mathcal{S}}  + (\nu,\rho,\lambda)_P\right] \,. \cr
\end{split}
\end{equation}
We explicitly show why the quantity $\dot{R}_{\mu,\text{SD}}$ depends on the shape of the wavepacket in the next subsection.
The first line describes the wavepacket's response to a quadratic potential through the quantum correlation of Berry curvature and position operator. 
The second line captures the response to a cubic potential via a more complicated quantum correlation.
Similarly, the third line corresponds to quantum correlation of the quantum metric dipole and the position operator. 
All of these terms are associated with the wavepacket's shape in both real ($\hat{r}$) and momentum space through the Berry curvature $\hat{\boldsymbol{\Omega}}(\mathbf{q})$ and the derivative of quantum metric $[ \hat{\boldsymbol{D}}, \hat{\boldsymbol{g}}(\mathbf{q})]$.

\subsection{Shape-dependent contributions}

We now compare our equations with those of Ref.~\cite{Kozii-PhysRevLett.126.156602} to show that our formalism yields the same results and explicitly reveals the shape-dependence contained in $\dot{R}_{\mu,\text{SD}}$. 
For a wavepacket
\begin{equation}
\ket{\Psi} = \mathcal{N}\sum_{n} \int \mathrm{d}^D \mathbf{q}~c_n(\mathbf{q}) \ket{\psi^{(n)}_{\mathbf{q}}} \,,
\end{equation}
its mean position is given by
\begin{equation}
\begin{split}
\braket{\hat{r}_\mu}_{\mathcal{S}} &= \sum_{n \in \mathcal{S}}\int \mathrm{d}^{D}\mathbf{q}~ |c_{n}(\mathbf{q})|^2 \partial_\mu \gamma_n(\mathbf{q}) + \sum_{m,n\in\mathcal{S}}\int \mathrm{d}^D \mathbf{q}~ c^\ast_{m}(\mathbf{q}) c_n(\mathbf{q}) A^{(m,n)}_\mu(\mathbf{q}) \,,
\end{split}
\end{equation}
where we defined $c_n(\mathbf{q}) = |c_n(\mathbf{q})|e^{-i \gamma_n(\mathbf{q})}$. 
In the Abelian case, we drop the band index and obtain :
\begin{equation}
\begin{split}
\braket{\hat{r}_\mu}_{\mathcal{S}} &= \int \mathrm{d}^{D}\mathbf{q}~ |c(\mathbf{q})|^2 \left[ \partial_\mu \gamma(\mathbf{q}) + A_\mu(\mathbf{q}) \right] = \int \mathrm{d}^{D}\mathbf{q}~ |c(\mathbf{q})|^2 \widetilde{R}_\mu \,,
\end{split}
\end{equation}
with $\widetilde{R}_\mu = \partial_\mu\gamma + A_\mu$. 
We can rewrite
\begin{equation}
\begin{split}
\partial_\mu \gamma &= \frac{i}{2|c(\mathbf{q})|^2} \left[ c^\ast(\mathbf{q}) \partial_\mu c(\mathbf{q}) - c(\mathbf{q}) \partial_\mu c^\ast (\mathbf{q})\right] \,.
\end{split}
\end{equation}
When we deal with $\left\langle \left\{ \hat{r}_{\nu,\mathcal{S}}, \boldsymbol{\Omega}_{\mu\rho} \right\} \right\rangle_{\mathcal{S}}$ in Eq.~\eqref{eq:RSD}, we obtain
\begin{equation}
\begin{split}
\frac{1}{2} \left\langle \left\{ \hat{r}_{\nu,\mathcal{S}}, \boldsymbol{\Omega}_{\mu\rho} \right\} \right\rangle_{\mathcal{S}} &= \frac{i}{2} \sum_{m,n\in\mathcal{S}}\int \mathrm{d}^{D}\mathbf{q}~\left[\partial_\nu c_m^\ast(\mathbf{q}) c_n(\mathbf{q}) - c_m^\ast(\mathbf{q}) \partial_\nu c_n(\mathbf{q})  \right] \Omega_{\mu\rho}^{(m,n)}(\mathbf{q}) \cr
&+ \frac{1}{2} \sum_{m,n\in\mathcal{S}}\int \mathrm{d}^{D}\mathbf{q}~ c_m^\ast(\mathbf{q}) c_n(\mathbf{q}) \left[ A_\nu^{(m,l)} \Omega_{\mu\rho}^{(l,n)}(\mathbf{q}) + \Omega_{\mu\rho}^{(m,l)}(\mathbf{q}) A_\nu^{(l,n)} \right] \,,
\end{split}
\end{equation}
and in the Abelian case, it reduces to
\begin{equation}
\begin{split}
\frac{1}{2} \left\langle \left\{ \hat{r}_\nu, \boldsymbol{\Omega}_{\mu\rho} \right\} \right\rangle_{\mathcal{S}} &= \int \mathrm{d}^{D}\mathbf{q}~|c(\mathbf{q})|^2\widetilde{R}_\nu \Omega_{\mu\rho}(\mathbf{q}) \,,
\end{split}
\end{equation}
where again $\widetilde{R}_\mu = \partial_\mu \gamma + A_\mu$. 
After a tedious calculation, the cubic contribution in Eq.~\eqref{eq:RSD} can also be expressed in a similar way, and yields
\begin{equation}
\begin{split}
\frac{1}{4} \left\langle \left\{ \hat{\boldsymbol{\Omega}}_{\mu\nu}, \left\{ \hat{r}_{\rho,\mathcal{S}}, \hat{r}_{\lambda,\mathcal{S}} \right\} \right\} \right\rangle_{\mathcal{S}} &= \int \mathrm{d}^D \mathbf{q}~ |c(\mathbf{q})|^2\Omega_{\mu\nu}(\mathbf{q}) \widetilde{R}_\rho \widetilde{R}_\lambda + \frac{1}{4|c(\mathbf{q})|^2}\int \mathrm{d}^D \mathbf{q}~ \Omega_{\mu\nu}(\mathbf{q}) \partial_\rho (|c(\mathbf{q})|^2) \partial_\lambda (|c(\mathbf{q})|^2) \,.
\end{split}
\end{equation}
The second term reveals a shape-dependent contribution in momentum space. Notice that Ref.~\cite{Kozii-PhysRevLett.126.156602}'s result corresponds to the Abelian case. In the non-Abelian framework, expressing shape-dependent terms using $c(\mathbf{q})$ becomes inelegant and increasingly cumbersome. Therefore, the operator-based projected Heisenberg equation offers a systematic and efficient approach for deriving semiclassical dynamics, especially for higher-order expansions.

As discussed in Ref.~\cite{Kozii-PhysRevLett.126.156602}, when we take the delta function limit for a narrow wavepacket in momentum space, namely when the shape of the wavepacket in momentum space becomes irrelevant, these terms cancel, meaning that the correlations in Eq.~\eqref{eq:RSD} vanish. 
This explain the fact that a non-vanishing contribution is related to the wavepacket shape, namely the coefficients $c(\mathbf q)$.
In our numerics, we however heuristically find that even when the wavepacket spread in momentum space, thus getting away from the delta function limit, the shape-dependent terms are still negligible and can be neglected.

\subsection{Semiclassical equations of wavepacket momentum velocity}

In calculating the evolution of the momentum operator, the dynamical behavior can be understood straightforwardly through the commutation relation between $\hat{q}^\mu$ and the projected position operators $\hat{r}_{\mu,\mathcal{S}}$, $\hat{r}^{(2)}_{\mu\nu,\mathcal{S}}$ and $\hat{r}^{(3)}_{\mu\nu\rho,\mathcal{S}}$
\begin{align}
[\hat{q}^\mu, \hat{r}_{\nu,\mathcal{S}}] &= -i\delta^{\mu}_{\nu} \,, \label{eq:[q,r]} \\
[\hat{q}^\mu, \hat{r}^{(2)}_{\nu\rho,\mathcal{S}}] &= -i\delta^{\mu}_{\nu} \hat{r}_{\rho,\mathcal{S}} - i\delta^{\mu}_{\rho} \hat{r}_{\nu,\mathcal{S}} \,, \label{eq:[q,rr]} \\
[\hat{q}^\mu, \hat{r}^{(3)}_{\nu\rho\lambda,\mathcal{S}}] &= - \frac{i}{2}\delta^{\mu}_{\nu} \left\{ \hat{r}_{\rho,\mathcal{S}}, \hat{r}_{\lambda,\mathcal{S}} \right\} - i \delta^{\mu}_{\nu} \hat{\boldsymbol{g}}_{\rho \lambda} + (\nu, \rho, \lambda)_P \,. \label{eq:[q,rrr]}
\end{align}
The velocity for the average momentum are given by
\begin{align}
\dot{Q}^\mu_0 &= 0 \,,\\
\dot{Q}^\mu_1 &= - E^\mu \,,\\
\dot{Q}^{\mu}_2 &= - \frac{1}{2} \left( E^{\mu\mu}\braket{ \hat{r}_\mu}_\mathcal{S} + E^{\mu\nu}\braket{ \hat{r}_\nu}_\mathcal{S}\right) \,,\\
\dot{Q}^{\mu}_3 &=- \frac{1}{2} E^{\mu\nu\rho} \left[ \frac{1}{2} \left\langle \left\{ \hat{r}_{\nu,\mathcal{S}}, \hat{r}_{\rho,\mathcal{S}} \right\} \right\rangle_{\mathcal{S}} + \braket{\hat{\boldsymbol{g}}_{\nu\rho}}_{\mathcal{S}} \right] = - \frac{1}{2} E^{\mu\nu\rho} \braket{\hat{r}_\nu \hat{r}_\rho}_{\mathcal{S}} \,,
\end{align}
and here we used Eq.~\eqref{eq:r3 matrix element} for $\dot{Q}^{\mu}_3 $.
These terms can be interpreted as a generalized form of Newton's second law, with corrections arising from quantum geometry and the internal structure (multipole moment) of the wavepacket. 
However, since the Brillouin zone manifold is a torus, this semiclassical formulation based on a well-defined $\hat{Q}$, remains valid only when the wavepacket in momentum space stays sufficiently localized and does not wrap around the entire Brillouin zone torus.

\subsection{Semiclassical equations of wavepacket real-space variance}

We have found numerical evidence that the dynamics of a Gaussian wavepacket in the presence of quadratic potential remains Gaussian over time. 
For this reason, it is worth deriving the EOM for the variance $W_{R,\mu\nu}$ in the case of quadratic potentials. 
For higher-order (such as cubic) potentials the variance is not anymore sufficient to describe the dynamics and a separate treatment is required.

To derive the equations of motion, we first compute the relevant commutators
\begin{align}
& \left[ \hat{r}^{(2)}_{\mu\nu,\mathcal{S}}, \hat{r}^{(2)}_{\rho\lambda, \mathcal{S}} \right] = \frac{i}{4} \left\{ \hat{r}_{\rho,\mathcal{S}}, \left\{ \hat{r}_{\mu,\mathcal{S}}, \hat{\boldsymbol{\Omega}}_{\nu\lambda} \right\} \right\} + \frac{i}{4} \left\{ \hat{r}_{\rho,\mathcal{S}}, \left\{ \hat{r}_{\nu,\mathcal{S}}, \hat{\boldsymbol{\Omega}}_{\mu\lambda} \right\} \right\} + \frac{i}{4} \left\{ \hat{r}_{\lambda,\mathcal{S}}, \left\{ \hat{r}_{\mu,\mathcal{S}}, \hat{\boldsymbol{\Omega}}_{\nu\rho} \right\} \right\} + \frac{i}{4} \left\{ \hat{r}_{\lambda,\mathcal{S}}, \left\{ \hat{r}_{\nu,\mathcal{S}}, \hat{\boldsymbol{\Omega}}_{\mu\rho} \right\} \right\} \nonumber \\
&+ \frac{i}{2} \left\{ \hat{r}_{\mu,\mathcal{S}}, \left[ \hat{\boldsymbol{D}}_\nu, \hat{\boldsymbol{g}}_{\rho\lambda} \right] \right\} + \frac{i}{2} \left\{ \hat{r}_{\nu,\mathcal{S}}, \left[ \hat{\boldsymbol{D}}_\mu, \hat{\boldsymbol{g}}_{\rho\lambda} \right] \right\} - \frac{i}{2} \left\{ \hat{r}_{\rho,\mathcal{S}}, \left[ \hat{\boldsymbol{D}}_\lambda, \hat{\boldsymbol{g}}_{\mu\nu} \right] \right\} - \frac{i}{2} \left\{ \hat{r}_{\lambda,\mathcal{S}}, \left[ \hat{\boldsymbol{D}}_\rho, \hat{\boldsymbol{g}}_{\mu\nu} \right] \right\} + \left[ \hat{\boldsymbol{g}}_{\mu\nu}, \hat{\boldsymbol{g}}_{\rho\lambda} \right] \,. 
\end{align}
By employing the Bloch basis representation, we calculate the variance velocity while retaining only the influence of the quadratic potential
\begin{align}
\frac{d}{dt} \braket{\hat{r}_\mu \hat{r}_\nu}_0 &= -i \left\langle \left[ \hat{r}^{(2)}_{\mu\nu}, \hat{H}_0 \right] \right\rangle_{\mathcal{S}} = \left[ \frac{1}{2} \left\langle \left\{ \hat{r}_{\mu,\mathcal{S}},[\hat{\boldsymbol{D}}_\nu, H_0] \right\} \right\rangle_{\mathcal{S}} + (\mu \leftrightarrow \nu) \right]  - i\left\langle \left[ \hat{\boldsymbol{g}}_{\mu\nu},H_0 \right] \right\rangle_{\mathcal{S}} \,, \\
\frac{d}{dt} \left\langle \hat{r}_\mu \hat{r}_\nu \right\rangle_1 &= - i E^\rho \left\langle \left[ \hat{r}^{(2)}_{\mu\nu}, \hat{r}_{\rho,\mathcal{S}} \right] \right\rangle_{\mathcal{S}} = E^\rho \left[ \frac{1}{2} \left\langle \left\{ \hat{r}_{\mu,\mathcal{S}},  \hat{\boldsymbol{\Omega}}_{\nu\rho} \right\} \right\rangle_{\mathcal{S}} + (\mu \leftrightarrow\nu) \right] - E^\rho \left\langle \left[ \hat{\boldsymbol{D}}_\rho, \hat{\boldsymbol{g}}_{\mu\nu} \right] \right\rangle_{\mathcal{S}} \,, \\ 
\frac{d}{dt} \braket{\hat{r}_\mu \hat{r}_\nu}_2 &= - \frac{i}{2} E^{\rho\lambda} \left\langle \left[ \hat{r}^{(2)}_{\mu\nu}, \hat{r}^{(2)}_{\rho\lambda} \right] \right\rangle_{\mathcal{S}} \nonumber \\
&= \frac{1}{8} E^{\rho\lambda} \left[ \left\langle \left\{ \hat{r}_{\mu,\mathcal{S}}, \left\{ \hat{r}_{\rho,\mathcal{S}}, \hat{\boldsymbol{\Omega}}_{\nu\lambda} \right\} \right\} \right\rangle_{\mathcal{S}} + (\mu \leftrightarrow \nu, \rho \leftrightarrow \lambda)\right] + \frac{1}{4} E^{\rho\lambda} \left[ \left\langle \left\{ \hat{r}_{\mu,\mathcal{S}}, \left[ \hat{\boldsymbol{D}}_\nu, \hat{\boldsymbol{g}}_{\rho\lambda} \right] \right\} \right\rangle_{\mathcal{S}} + (\mu \leftrightarrow \nu)\right] \nonumber \\
& - \frac{1}{4} E^{\rho\lambda} \left[ \left\langle \left\{ \hat{r}_{\rho,\mathcal{S}}, \left[ \hat{\boldsymbol{D}}_\lambda, \hat{\boldsymbol{g}}_{\mu\nu} \right] \right\} \right\rangle_{\mathcal{S}} + (\rho \leftrightarrow \lambda)\right] - \frac{i}{2} E^{\rho\lambda} \left\langle \left[ \hat{\boldsymbol{g}}_{\mu\nu}, \hat{\boldsymbol{g}}_{\rho\lambda} \right] \right\rangle_{\mathcal{S}} \,,
\end{align}
and
\begin{align}
\braket{\hat{r}_\nu}_{\mathcal{S}} \frac{d}{dt} \braket{\hat{r}_\mu}_0 &= \braket{ \hat{r}_\nu}_{\mathcal{S}} \left\langle \left[ \hat{\boldsymbol{D}}_\mu, \hat{H}_0 \right] \right\rangle_{\mathcal{S}} \,, \\
\braket{\hat{r}_\nu}_{\mathcal{S}} \frac{d}{dt} \braket{\hat{r}_\mu}_1 &= E^\rho \braket{ \hat{r}_\nu}_{\mathcal{S}} \braket{\hat{\boldsymbol{\Omega}}_{\mu\rho}}_{\mathcal{S}} \,, \\
\braket{\hat{r}_\nu}_{\mathcal{S}} \frac{d}{dt} \braket{\hat{r}_\mu}_2 &= \frac{1}{4} E^{\rho\lambda} \braket{ \hat{r}_\nu}_{\mathcal{S}} \left[ \left\langle \left\{ \hat{r}_{\rho,\mathcal{S}}, \hat{\boldsymbol{\Omega}}_{\mu\lambda} \right\} \right\rangle_{\mathcal{S}} + (\rho \leftrightarrow \lambda) \right] + \frac{1}{2} E^{\rho\lambda} \braket{ \hat{r}_\nu}_{\mathcal{S}} \left\langle \left[ \hat{\boldsymbol{D}}_\mu, \hat{\boldsymbol{g}}_{\rho\lambda} \right] \right\rangle_{\mathcal{S}} \,.
\end{align}
Finally, the equations of $W_{R,\mu\nu}$ are given by
\begin{align}
[\dot{W}_{R,\mu\nu}]_0 &=  \left[ \frac{1}{2} \left\langle \left\{ \hat{r}_{\mu,\mathcal{S}},[\hat{\boldsymbol{D}}_\nu, H_0] \right\} \right\rangle_{\mathcal{S}} - \braket{ \hat{r}_\mu}_{\mathcal{S}} \left\langle \left[ \hat{\boldsymbol{D}}_\nu, \hat{H}_0 \right] \right\rangle_{\mathcal{S}} + (\mu \leftrightarrow \nu) \right]  - i\left\langle \left[ \hat{\boldsymbol{g}}_{\mu\nu},H_0 \right] \right\rangle_{\mathcal{S}} \,, \label{eq:dVarR0}\\
[\dot{W}_{R,\mu\nu} ]_1 &=  E^\rho \left[ \frac{1}{2} \left\langle \left\{ \hat{r}_{\mu,\mathcal{S}},  \hat{\boldsymbol{\Omega}}_{\nu\rho} \right\} \right\rangle_{\mathcal{S}} - \braket{ \hat{r}_\mu}_{\mathcal{S}} \braket{\hat{\boldsymbol{\Omega}}_{\nu\rho}}_{\mathcal{S}} + (\mu \leftrightarrow\nu) \right] - E^\rho \left\langle \left[ \hat{\boldsymbol{D}}_\rho, \hat{\boldsymbol{g}}_{\mu\nu} \right] \right\rangle_{\mathcal{S}} \,,  \label{eq:dVarR1} \\
[\dot{W}_{R,\mu\nu} ]_2 &= \frac{1}{2} E^{\rho\lambda} \left[ \frac{1}{4} \left\langle \left\{ \hat{r}_{\mu,\mathcal{S}}, \left\{ \hat{r}_{\rho,\mathcal{S}}, \hat{\boldsymbol{\Omega}}_{\nu\lambda} \right\} \right\} \right\rangle_{\mathcal{S}} - \frac{1}{2} \braket{ \hat{r}_\mu}_{\mathcal{S}} \left\langle \left\{ \hat{r}_{\rho,\mathcal{S}}, \hat{\boldsymbol{\Omega}}_{\nu\lambda} \right\} \right\rangle_{\mathcal{S}} + (\mu \leftrightarrow \nu, \rho \leftrightarrow \lambda)\right] \nonumber \\
&+ \frac{1}{2} E^{\rho\lambda} \left[ \frac{1}{2} \left\langle \left\{ \hat{r}_{\mu,\mathcal{S}}, \left[ \hat{\boldsymbol{D}}_\nu, \hat{\boldsymbol{g}}_{\rho\lambda} \right] \right\} \right\rangle_{\mathcal{S}} - \braket{\hat{r}_{\mu}}_{\mathcal{S}} \left\langle \left[ \hat{\boldsymbol{D}}_\nu, \hat{\boldsymbol{g}}_{\rho\lambda} \right] \right\rangle_{\mathcal{S}} + (\mu \leftrightarrow \nu)\right] \nonumber \\
&- \frac{1}{4} E^{\rho\lambda} \left[ \left\{ \hat{r}_{\rho,\mathcal{S}}, \left[ \hat{\boldsymbol{D}}_\lambda, \hat{\boldsymbol{g}}_{\mu\nu} \right] \right\} + (\rho \leftrightarrow \lambda) \right] - \frac{i}{2} E^{\rho\lambda} \left\langle \left[ \hat{\boldsymbol{g}}_{\mu\nu}, \hat{\boldsymbol{g}}_{\rho\lambda} \right] \right\rangle_{\mathcal{S}} \,.
 \label{eq:dVarR2}
\end{align}
The first term of Eq.~\eqref{eq:dVarR0} accounts for the contribution from band dispersion and has also been studied in Ref.~\cite{Souza-PhysRevB.79.045127}. 
The intrinsic fluctuation term $[\dot{W}_{R,\mu\nu}]_0$ arises from the different dynamical phase accumulations in the Bloch states $\ket{\psi_{\mathbf{q}}}$, leading to gradual changes in the wavepacket's variance over time.
In the case of a degenerate flat band, Eq.~\eqref{eq:dVarR0} vanishes identically.
Eq.~\eqref{eq:dVarR1}, which originates from an external linear potential, contains a first term that depends on the wavepacket shape. 
This term vanishes in one-dimensional systems or in models possessing $\mathcal{PT}$ symmetry.
Finally, Eq.~\eqref{eq:dVarR2} also reflects a wavepacket shape-dependent contribution to the variance velocity.\\
In the main text, we decompose these equations into shape-independent
\begin{equation}
\begin{split}
(\dot{W}_{R,\mu\nu})_{\text{SI}} &= - i\left\langle \left[ \hat{\boldsymbol{g}}_{\mu\nu},H_0 \right] \right\rangle_{\mathcal{S}} - E^\rho(\mathbf{R}) \left\langle \left[ \hat{\boldsymbol{D}}_\rho, \hat{\boldsymbol{g}}_{\mu\nu} \right] \right\rangle_{\mathcal{S}}  - \frac{i}{2} E^{\rho\lambda} \left\langle \left[ \hat{\boldsymbol{g}}_{\mu\nu}, \hat{\boldsymbol{g}}_{\rho\lambda} \right] \right\rangle_{\mathcal{S}} \,, \cr
\end{split}
\end{equation}
where $E^\rho(\mathbf{R}) = E^\rho + \frac{1}{2} \left( E^{\rho\rho} \braket{\hat{r}_\rho}_{\mathcal{S}} + E^{\rho\lambda} \braket{ \hat{r}_\lambda }_{\mathcal{S}} \right)$ and shape-dependent parts
\begin{equation}
\begin{split}
(\dot{W}_{R,\mu\nu})_{\text{SD}} &=  \left[ \frac{1}{2} \left\langle \left\{ \hat{r}_{\mu,\mathcal{S}},[\hat{\boldsymbol{D}}_\nu, H_0] \right\} \right\rangle_{\mathcal{S}} - \braket{ \hat{r}_\mu}_{\mathcal{S}} \left\langle \left[ \hat{\boldsymbol{D}}_\nu, \hat{H}_0 \right] \right\rangle_{\mathcal{S}} + (\mu \leftrightarrow \nu) \right] \cr
&+ E^\rho \left[ \frac{1}{2} \left\langle \left\{ \hat{r}_{\mu,\mathcal{S}},  \hat{\boldsymbol{\Omega}}_{\nu\rho} \right\} \right\rangle_{\mathcal{S}} - \braket{ \hat{r}_\mu}_{\mathcal{S}} \braket{\hat{\boldsymbol{\Omega}}_{\nu\rho}}_{\mathcal{S}} + (\mu \leftrightarrow\nu) \right] \cr
&+ \frac{1}{2} E^{\rho\lambda} \left[ \frac{1}{4} \left\langle \left\{ \hat{r}_{\mu,\mathcal{S}}, \left\{ \hat{r}_{\rho,\mathcal{S}}, \hat{\boldsymbol{\Omega}}_{\nu\lambda} \right\} \right\} \right\rangle_{\mathcal{S}} - \frac{1}{2} \braket{ \hat{r}_\mu}_{\mathcal{S}} \left\langle \left\{ \hat{r}_{\rho,\mathcal{S}}, \hat{\boldsymbol{\Omega}}_{\nu\lambda} \right\} \right\rangle_{\mathcal{S}} + (\mu \leftrightarrow \nu, \rho \leftrightarrow \lambda)\right] \cr
&+ \frac{1}{2} E^{\rho\lambda} \left[ \frac{1}{2} \left\langle \left\{ \hat{r}_{\mu,\mathcal{S}}, \left[ \hat{\boldsymbol{D}}_\nu, \hat{\boldsymbol{g}}_{\rho\lambda} \right] \right\} \right\rangle_{\mathcal{S}} - \braket{\hat{r}_{\mu}}_{\mathcal{S}} \left\langle \left[ \hat{\boldsymbol{D}}_\nu, \hat{\boldsymbol{g}}_{\rho\lambda} \right] \right\rangle_{\mathcal{S}} + (\mu \leftrightarrow \nu)\right] \cr
&- \frac{1}{2} E^{\rho\lambda} \left[ \frac{1}{2} \left\langle \left\{ \hat{r}_{\rho,\mathcal{S}}, \left[ \hat{\boldsymbol{D}}_\lambda, \hat{\boldsymbol{g}}_{\mu\nu} \right] \right\} \right\rangle_{\mathcal{S}} - \braket{\hat{r}_\rho}_{\mathcal{S}} \left\langle \left[ \hat{\boldsymbol{D}}_\lambda, \hat{\boldsymbol{g}}_{\mu\nu} \right] \right\rangle_{\mathcal{S}} + (\rho \leftrightarrow \lambda) \right] \,.
\end{split}
\end{equation}
Notice that in the one-dimensional flat band case, all these terms are exactly zero, namely $ (\dot{W}_{R,xx})_{\text{SD}} = 0$.

\subsection{Semiclassical equations of wavepacket momentum variance: dispersion in momentum space }

The dispersion behavior of a wavepacket in momentum space is characterized by its variance, defined as $W_{Q}^{\mu\nu} = \braket{\hat{q}^\mu \hat{q}^\nu}_{\mathcal{S}} - \braket{\hat{q}^\mu}_{\mathcal{S}} \braket{\hat{q}^\nu}_{\mathcal{S}}$. As a first step in evaluating the time evolution of this quantity, we compute the relevant commutators
\begin{align}
[\hat{q}^\mu \hat{q}^\nu, \hat{r}_{\rho,\mathcal{S}}] &= -i \delta^{\mu}_{\rho} \hat{q}^\nu - i \delta^{\nu}_{\rho} \hat{q}^\mu \,, \\
[\hat{q}^\mu \hat{q}^\nu, \hat{r}^{(2)}_{\rho\lambda, \mathcal{S}}] &= -\frac{i}{2} \left[ \delta^{\mu}_{\lambda} \{\hat{q}^\nu, \hat{r}_{\rho,\mathcal{S}} \} + (\rho \leftrightarrow \lambda)  \right] -\frac{i}{2} \left[ \delta^{\nu}_{\lambda} \{\hat{q}^\mu, \hat{r}_{\rho,\mathcal{S}} \} + (\rho \leftrightarrow \lambda)  \right] \,.
\label{eq:qqr}
\end{align}
Following the definition of $W_{Q}^{\mu\nu}$, we obtain
\begin{align}
[\dot{W}_{Q}^{\mu\nu}]_0 &= [\dot{W}_{Q}^{\mu\nu}]_1 = 0 \,, \\
[\dot{W}_{Q}^{\mu\nu}]_2 &= -\frac{E^{\rho\lambda}}{2} \left[ \delta^{\mu}_{\lambda} \left( \frac{1}{2} \left\langle \left\{ \hat{q}^\nu,\hat{r}_{\rho} \right\} \right\rangle_{\mathcal{S}} - \braket{\hat{q}^\nu}_{\mathcal{S}}  \braket{\hat{r}_\rho}_{\mathcal{S}} \right) + (\rho \leftrightarrow \lambda) \right] + ( \mu \leftrightarrow \nu) \nonumber \\
&= -E^{\mu\rho} W_{RQ,\rho}^{\nu} + ( \mu \leftrightarrow \nu) \,. \label{eq:dVarQ2}
\end{align}
This indicates that $\dot{W}_{Q}^{\mu\nu}$ is related to the phase space variance $W_{RQ,\rho}^{\nu}\equiv \frac{1}{2} \left\langle \left\{ \hat{q}^\nu,\hat{r}_{\rho} \right\} \right\rangle_{\mathcal{S}} - \braket{\hat{q}^\nu}_{\mathcal{S}}  \braket{\hat{r}_\rho}_{\mathcal{S}} = \left\langle  \hat{q}^\nu \hat{r}_{\rho}\right\rangle_{\mathcal{S}} - \braket{\hat{q}^\nu}_{\mathcal{S}}  \braket{\hat{r}_\rho}_{\mathcal{S}}$. 
However, this relationship alone is not sufficient to fully determine the wavepacket's dispersion dynamics. 
Consequently, it is necessary to derive the equations governing the phase space variance $W_{RQ,\rho}^{\nu}$.

\subsection{ Semiclassical equations of wavepacket phase space variance}

Since $W_{RQ,\mu}^{\nu}$ is time-dependent due to the evolution of $\mathbf{R}$ and $\mathbf{Q}$, we derive the corresponding Heisenberg equations:
\begin{align}
\dot{W}_{RQ,\mu}^{\nu} &=  \frac{1}{2} \frac{d}{dt} \left\langle \left\{\hat{r}_{\mu,\mathcal{S}}, \hat{q}^\nu \right\} \right\rangle_{\mathcal{S}} - \dot{R}_{\mu} Q^\nu - R_\mu \dot{Q}^\nu \,. \label{eq:dVarRQ2}
\end{align}
Here we evaluate commutators using of Eqs.~\eqref{eq:[r,r]}-\eqref{eq:[r,rr]} and Eqs.\eqref{eq:[q,r]}-\eqref{eq:[q,rr]}:
\begin{align}
\frac{1}{2} \left[ \left\{ \hat{r}_{\mu,\mathcal{S}}, \hat{q}^\nu \right\}, \hat{H}_0 \right] &= - \frac{1}{2} \left\{ \left[ \hat{H}_0, \hat{q}^\nu \right], \hat{r}_{\mu,\mathcal{S}} \right\} - \frac{1}{2} \left\{ \left[ \hat{H}_0, \hat{r}_{\mu,\mathcal{S}} \right], \hat{q}^\nu \right\}= \frac{i}{2} \left\{ \left[ \hat{\boldsymbol{D}}_\mu, \hat{H}_0 \right], \hat{q}^\nu \right\} \,, \\
\frac{1}{2} \left[ \left\{ \hat{r}_{\mu,\mathcal{S}}, \hat{q}^\nu \right\}, \hat{r}_{\rho,\mathcal{S}} \right] &= -\frac{1}{2} \left\{ \left[ \hat{r}_{\rho,\mathcal{S}}, \hat{q}^\nu \right], \hat{r}_{\mu,\mathcal{S}} \right\} - \frac{1}{2} \left\{ \left[ \hat{r}_{\rho,\mathcal{S}}, \hat{r}_{\mu,\mathcal{S}} \right], \hat{q}^\nu \right\}= -i \delta_{\rho}^{\nu} \hat{r}_{\mu,\mathcal{S}} + \frac{i}{2} \left\{ \hat{\boldsymbol{\Omega}}_{\mu\rho},\hat{q}^\nu \right\} \,, \\
\frac{1}{2} \left[ \left\{ \hat{r}_{\mu,\mathcal{S}}, \hat{q}^\nu \right\}, \hat{r}^{(2)}_{\rho\lambda,\mathcal{S}} \right] &= -\frac{1}{2} \left\{ \left[ \hat{r}^{(2)}_{\rho\lambda, \mathcal{S} }, \hat{q}^\nu \right], \hat{r}_{\mu,\mathcal{S}} \right\} - \frac{1}{2} \left\{ \left[ \hat{r}^{(2)}_{\rho\lambda, \mathcal{S} }, \hat{r}_{\mu,\mathcal{S}} \right], \hat{q}^\nu \right\} \nonumber \\
= -\frac{i}{2} \delta^{\nu}_{\rho} \left\{ \hat{r}_{\lambda,\mathcal{S}}, \hat{r}_{\mu,\mathcal{S}} \right\} &- \frac{i}{2} \delta^{\nu}_{\lambda} \left\{ \hat{r}_{\rho,\mathcal{S}}, \hat{r}_{\mu,\mathcal{S}} \right\} + \frac{i}{4} \left\{ \left\{ \hat{r}_{\rho,\mathcal{S}}, \hat{\boldsymbol{\Omega}}_{\mu\lambda} \right\}, \hat{q}^\nu \right\} + \frac{i}{4} \left\{ \left\{ \hat{r}_{\lambda,\mathcal{S}}, \hat{\boldsymbol{\Omega}}_{\mu\rho} \right\}, \hat{q}^\nu \right\} + \frac{i}{2} \left\{ \left[ \hat{\boldsymbol{D}}_\mu, \hat{\boldsymbol{g}}_{\rho\lambda} \right] , \hat{q}^\nu \right\} \,.
\end{align}
After tedious calculations, we obtain
\begin{align}
[\dot{W}_{RQ,\mu}^{\nu}]_0 &= \frac{1}{2} \left\langle \left\{ \left[ \hat{\boldsymbol{D}}_\mu, \hat{H}_0 \right], \hat{q}^\nu \right\} \right\rangle_{\mathcal{S}} - \left\langle \left[ \hat{\boldsymbol{D}}_\mu, \hat{H}_0 \right] \right\rangle_{\mathcal{S}} \braket{\hat{q}^\nu}_{\mathcal{S}} \label{eq: VarRQ0} \,,\\
[\dot{W}_{RQ,\mu}^{\nu}]_1 &= E^\rho \left( \frac{1}{2} \left\langle \left\{ \hat{q}^\nu, \hat{\boldsymbol{\Omega}}_{\mu\rho} \right\} \right\rangle_{\mathcal{S}} - \braket{\hat{q}^\nu}_{\mathcal{S}} \left\langle \hat{\boldsymbol{\Omega}}_{\mu\rho} \right\rangle_{\mathcal{S}} \right) \label{eq: VarRQ1} \,, \\
[\dot{W}_{RQ,\mu}^{\nu}]_2 &= \frac{1}{2} E^{\rho\lambda} \left[ \frac{1}{4}\left\langle \left\{ \hat{q}^\nu, \left\{ \hat{r}_{\rho,\mathcal{S}}, \hat{\boldsymbol{\Omega}}_{\mu\lambda} \right\} \right\} \right\rangle_\mathcal{S} - \frac{1}{2} \braket{\hat{q}^\nu} \left\langle \left\{ \hat{r}_{\rho,\mathcal{S}}, \hat{\boldsymbol{\Omega}}_{\mu\lambda} \right \} \right\rangle_{\mathcal{S}} + (\rho \leftrightarrow \lambda) \right] \nonumber \\
&+ \frac{1}{2} E^{\rho\lambda} \left[ \frac{1}{2} \left\langle \left\{ \hat{q}^\nu, \left[ \hat{\boldsymbol{D}}_\mu, \hat{\boldsymbol{g}}_{\rho\lambda} \right] \right\} \right\rangle_{\mathcal{S}} - \braket{\hat{q}^\nu}_{\mathcal{S}} \left\langle \left[ \hat{\boldsymbol{D}}_\mu, \hat{\boldsymbol{g}}_{\rho\lambda} \right] \right\rangle_{\mathcal{S}} \right] \nonumber \\
&- E^{\nu\rho} \left( W_{R,\mu\rho} - \braket{ \hat{\boldsymbol{g}}_{\mu\rho}}_{\mathcal{S}} \right) \label{eq: VarRQ2} \,.
\end{align}
All the quantum correlator terms we have found to be negligible, analogously for the case of narrow wavepacket discussed above. 
The last term contains the real-space width and the quantum metric. 
The former is much larger than one (in units of the lattice spacing) and the latter is smaller than one in our models as we have a large gap between the flat bands and all the other bands. 
Therefore, as a rough approximation, we obtain the simple expression
\begin{equation}
\dot{W}_{RQ,\mu}^{\nu} \approx -E^{\nu\rho} W_{R,\mu\rho} \label{eq:VarRQ approx} \,.
\end{equation}

\section{Momentum space variance: approximate dynamics}

\begin{figure}[!t]
\centering
\includegraphics[width=16cm]{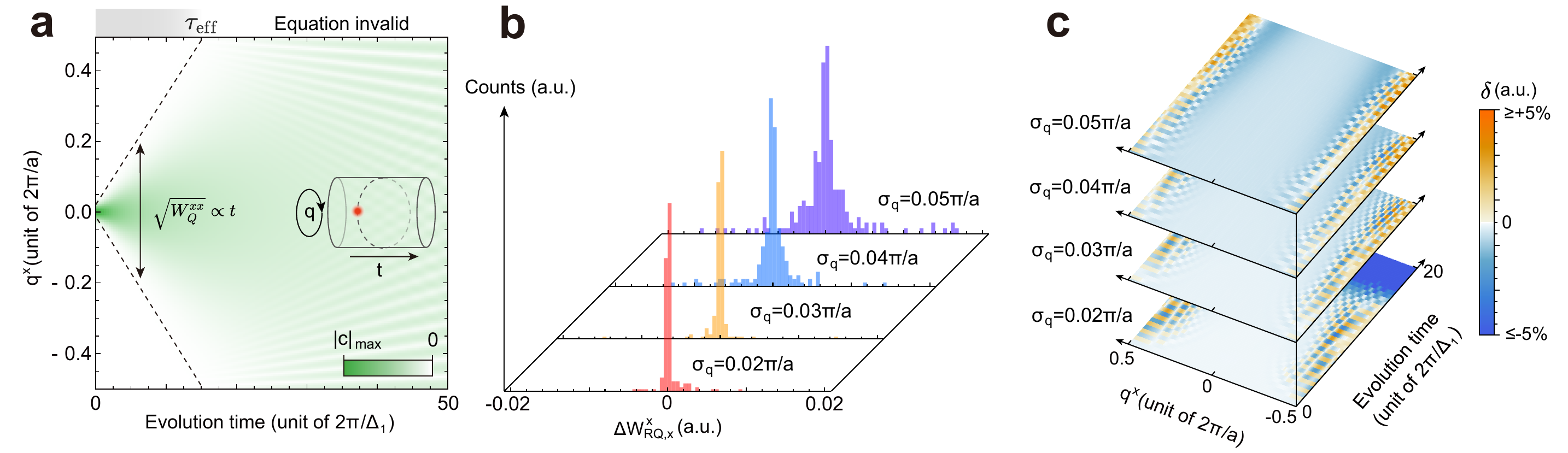}
\caption{ (a) Dispersion of the momentum space wavepacket under an external quadratic potential with strength $E^{xx}/\Delta_{1} = 10^{-3}$. 
(b) Fixing a single width $\sigma_q$, we prepare wavepackets for the whole momenta range $Q^{x}$ from $[0, 2\pi/a)$ and sample uniformly the phase-space variance 
$W_{RQ,x}^{x}(0)$ for 200 wavepackets.
(c) Deviation $\delta(t)$ from the approximation prediction. Other parameters are identical to those used in Fig.~3 of the main text. }
\label{fig:1d Lieb dispersion}
\end{figure}

In this section, we discuss the effective behaviour of the momentum space and phase space variances over time, for which several approximations can be done.
In particular, this will allow to quantify when the broadening of the wavepacket in momentum space will cause the wavepacket to wrap around the Brillouin zone and thus self-interfere, which corresponds to the breakdown of momentum space locality. 
Such regime determines the end of validity of the semiclassical equations for the momentum space variance. 
In turn this means that we closed a closed set of equations for the semiclassical dynamics. 

By combining Eq.~\eqref{eq:dVarQ2} and Eq.~\eqref{eq:VarRQ approx}, i.e. we neglect quantum correlator or shape dependent terms, we obtain the dynamics of the momentum space variance in the one-dimensional case:
\begin{equation}
\Delta W_Q^{xx} = 2 E^{xx} W_{RQ,x}^{x}(0) t + (E^{xx})^2 W_{R,xx}(0) t^2 = \alpha t + \beta t^2\,,
\label{eq:WQapprox}
\end{equation}
under the assumption that $W_{R,xx}(t)\approx W_{R,xx}(0)$.
In Fig.~\ref{fig:1d Lieb dispersion}(a), we present the dynamics of the momentum space variance, which indicates that $\alpha\approx 0$ or, in other words, that $\sqrt{W_{Q}^{xx}} \sim t$. 
This simple relation allows us to determine the time $\tau_{\mathrm{eff}}$ at which a Gaussian wavepacket will wrap around the entire Brillouin zone, which gives us a good estimate for the breakdown of the assumption of locality in momentum space.
This breakdown is reflected in the significant deviations in Fig.~3(e,f) insets of the main text and in Fig.~\ref{fig:1d Lieb dispersion}(c).

We present the numerical results of phase space variance $W_{RQ,x}^{x}(0)$ in Fig.~\ref{fig:1d Lieb dispersion}(b), obtained by preparing wavepacket with different momentum space standard deviations ($\sigma_q$) and different mean momenta $Q^{x}$ ranging from 0 to $2\pi/a$. 
We observe that in all these cases, which are relevant to this manuscript, $W_{RQ,x}^{x}(0)<10^{-2}$.
This allows us to find a condition for the importance of the linear term ($\alpha t$) as compared to the quadratic term ($\beta t^2$). 
We define a characteristic timescale $\tau_s = \alpha/\beta = W_{RQ,x}^{x}(0)/ \left[ E^{xx} W_{R,xx}(0) \right]$.
Since in our simulations $E^{xx}\approx 10^{-3}$ and $W_{R,xx}(0)\approx 10^2$ for a large wavepacket, we obtain $\tau_s \approx 10^{-2}$. 
As a result, for the dynamics shown in the figures, where we typically have $t\gtrsim 1$, we are deep in the condition of $t\gg \tau_s$ and this explains why we can neglect the $\alpha$ term.
As a result, the phase space variance is actually irrelevant and can be discarded safely. 
Since estimating the phase-space variance $W_{RQ,x}^{x}$ in experiments is not straightforward, having the opportunity to neglect as we do here allows have a closed set of equations that can be directly connected to the experimental observables.
In the end, this amounts to observing the linear scaling behavior $\sqrt{W_{Q}^{xx}} \sim t$, as shown in Fig.~\ref{fig:1d Lieb dispersion}(a), which is what allows us to easily estimate the breakdown of locality, as commented above.

To further convince ourselves of the correctness of Eq.~\eqref{eq:WQapprox}, we need to rule out the contribution of shape dependent terms. 
We thus consider the exact equation containing shape-dependent terms by combining Eq.~\eqref{eq:dVarQ2} and \eqref{eq: VarRQ2}.
We obtain:
\begin{equation}
\ddot{W}_{Q}^{xx} = \frac{1}{2} E^{xx} \left[ \frac{1}{2} \left\langle \left\{ \hat{q}^x, \partial_x g_{xx} \right\} \right\rangle_{\mathcal{S}} - \braket{\hat{q}^x}_{\mathcal{S}} \left\langle \partial_x g_{xx} \right\rangle_{\mathcal{S}} \right] + 2(E^{xx} )^2 \left( W_{R,xx} - g_{xx}\right)\,. \label{eq:1D Lieb ddVarQ}
\end{equation}
The real space variance on the right-hand side of this equation remains effectively constant over time as discussed before.
This is because, in our calculations, the real space variance is generally large and exhibits only minimal variation over time (as demonstrated in Fig.~3 of the main text).
Therefore, it is reasonable to approximate $ W_{R,xx} \approx W_{R,xx}(0)$.
The impact of the additional terms in Eq.~\eqref{eq:1D Lieb ddVarQ} can be estimated by defining
\begin{equation}
\delta(t) \equiv \frac{\frac{d^2}{dt^2}W_{Q}^{xx} - 2(E^{xx} )^2 W_{R,xx}(0)}{2(E^{xx} )^2 W_{R,xx}(0)}\,,
\end{equation}
which quantifies the relative deviation from the leading-order prediction $2(E^{xx})W_{R,xx}|_{t=0}$ during the actual evolution.
As shown in Fig.~\ref{fig:1d Lieb dispersion}(c), we plot the $\delta(t)$ for wavepackets with varying initial momentum and variance. 
The results demonstrate that the deviation remains within $\pm 5\%$, confirming the robustness and accuracy of the approximation.

\section{Numerical calculation in the non-Abelian case: One dimensional extended Lieb model}

\begin{figure}[t!]
\centering
\includegraphics[width=12cm]{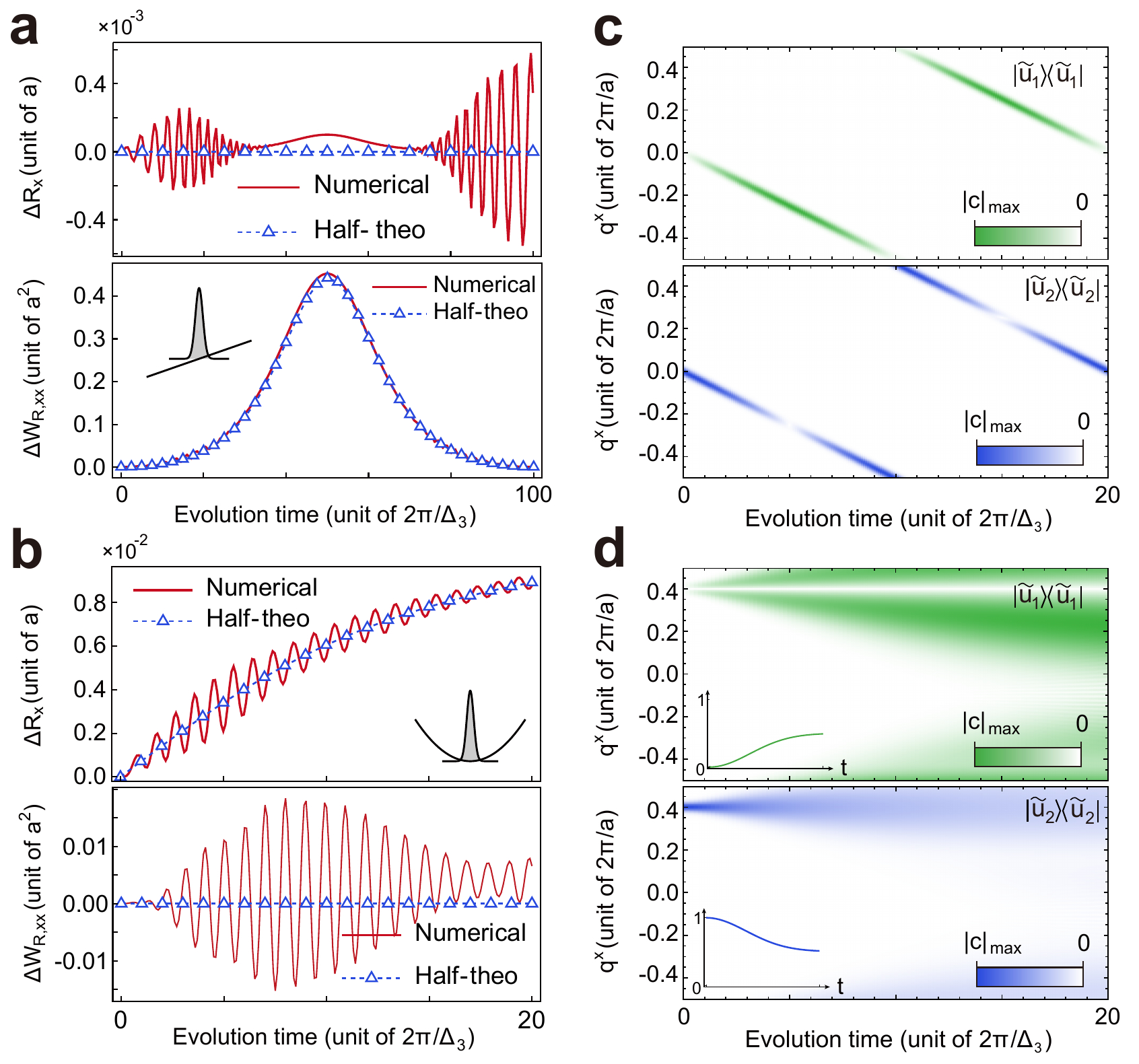}
\caption{ Time evolution of the wavepacket's center $R_x$ and spatial variance $W_{R,xx}$ under external (a) linear potential with strength $E^x/\Delta_3 = 10^{-2}$ and (b) quadratic potential with strength $E^{xx}/\Delta_3 = 10^{-3}$. 
(c,d) Corresponding momentum space distributions for the degenerate flat band states associated with the cases in (a) and (b), respectively. 
Other parameters are fixed as $t_1 = t_2 = t_3 = t_4 = \Delta_3/\sqrt{2}$ and the system size is set to $L=200$. }
\label{fig:1d extended Lieb trajectory}
\end{figure}

In this manuscript, we have introduced equations for non-Abelian models but in the main text we have now shown their validation. 
We fill this gap here by considering a non-Abelian tight-binding model with two degenerate flat bands that can be constructed by introducing an additional site, denoted $\mathrm{D}$, into the one-dimensional Lieb lattice (see Fig.~2 in the main text). 
This site connects only to site $\mathrm{A}$, as illustrated in Fig.~2. 
The corresponding Hamiltonian for this model, as proposed in Ref.~\cite{Brosco-PhysRevA.103.063518}, is given by
\begin{equation}
H_{3} = \sum_j (t_1 b^\dag_j a_j + t_2 a_{j+1}^\dag b_j + t_3 c_j^\dag a_j + t_4 d_j^\dag a_j) + \text{H.c.}.
\end{equation}
The system features an energy gap $\Delta_{3} = \sqrt{t_1^2 + t_2^2 + t_3^2+t_4^2 + 2 t_1 t_2 \cos{(q^x a)}}$. 
The eigenstates corresponding to the degenerate flat bands are given by:
\begin{equation}
\begin{split} 
\ket{u_{\mathrm{FB},1}(q^x)} &= \frac{1}{\delta_3} (0,0,-t_3, t_4)^T  \cr
\ket{u_{\mathrm{FB,2}}(q^x)} &= \frac{1}{\delta_{3} \Delta_{3}(q^x)} \left( 0, \delta_3^2, (t_1 + t_2 e^{i q^x})t_3, (t_1 + t_2 e^{i q^x}) t_4 \right)^T,
\end{split}
\end{equation}
where $\delta_{3} = \sqrt{t_3^2+t_4^2}$. In this basis, the state $\ket{u_{\mathrm{FB},1}}$ is momentum-independent, resulting in a vanishing inter-band Berry connection $A^{(21)} = i\braket{u_{\mathrm{FB},2} | \partial_\mu u_{\mathrm{FB},1}} = 0$. 
As a result, even under a linear potential, there is no inter-band mixing over time. 
Similarly, the off-diagonal components of the quantum metric vanish, indicating the absence of non-trivial non-Abelian effects in this model. 
This result is not surprising, given that we are dealing with a one-dimensional model, where non-trivial topological structure does not arise.
Consequently, it is always possible to perform a continuous gauge transformation that diagonalizes the Berry connection.

While this is a trivial non-Abelian model, we can still employ it to test the non-Abelian equations that we have introduced in the main text and in the previous sections.
Any state in the degenerate flat bands can be expressed as a
linear combination of the two basis states $\{ \ket{u_{\mathrm{FB},1}}, \ket{u_{\mathrm{FB},2}} \}$. 
As an illustrative example, we consider the following momentum dependent basis transformation:
\begin{equation}
\begin{split}
\ket{\widetilde{u}_1(q^x)} &= \cos{q^x} \ket{u_{\mathrm{FB},1}(q^x)} + \sin{q^x} \ket{u_{\mathrm{FB},2}(q^x)} \,, \cr
\ket{\widetilde{u}_2(q^x)} &= -\sin{q^x} \ket{u_{\mathrm{FB},1}(q^x)} + \cos{q^x} \ket{u_{\mathrm{FB},2}(q^x)} \,.
\end{split}
\end{equation} 
This new basis choice introduces non-zero off-diagonal components in both the Berry connection and the quantum metric.
The semiclassical EOM in the non-Abelian case, under the external potential $V_{\mathrm{ext}} = E^x x + \frac{1}{2} E^{xx}x^2$ are given by
\begin{align}
\dot{R}^{x} &= \frac{1}{2}E^{xx} \left\langle \left[ \hat{\boldsymbol{D}}_x, \tilde{\boldsymbol{g}}^{xx} \right] \right\rangle_{\mathcal{S}} \,, 
\label{eq:1D extended Lieb R} \\
\dot{Q}^{x} &= -E^{x}(R_x) \,, 
\label{eq:1D extended Lieb Q} \\
\dot{W}_{R,xx} &= -E^{x} \left\langle \left[ \hat{\boldsymbol{D}}_x, \tilde{\boldsymbol{g}}^{xx} \right] \right\rangle_{\mathcal{S}} \,,
\label{eq:1D extended Lieb W} \\
\dot{W}_{Q}^{xx} &= 2(E^{xx})^2 W_{R,xx}(0) t \,.
\label{eq:1D extended Lieb WQ}
\end{align}
The symbol $\widetilde{O}$ denotes the non-Abelian multi-band tensor defined in the new basis $\ket{\widetilde{u}_{1,2}}$.
The validity of Eqs.~\eqref{eq:1D extended Lieb R} and \eqref{eq:1D extended Lieb W} is confirmed by numerical simulations.
The resulting trajectories of $R_{x}$, $W_{R,xx}$ and the band populations are displayed in Fig.~\ref{fig:1d extended Lieb trajectory}.
The oscillation reflect coherent mixing dynamics arising from the gauge transformation applied to the degenerate flat band basis.
Notably, the displacement in Fig.~\ref{fig:1d extended Lieb trajectory}(a) originates from non-adiabatic effects.
Although the population remains close to unity in the flat band, the absence of dispersive dynamics in this band amplifies the relative impact of non-adiabatic transitions. 
Nevertheless, the resulting displacement remains much smaller than one lattice constant, indicating the robustness of the flat-band localization even under basis rotation.

\section{Constructing maximally localized wavepackets}

In general, a wavepacket can be expressed as $\ket{\Psi(\mathbf{r})} = \sum_\mathbf{q} c_n(\mathbf{q}) \ket{\psi^{(n)}_{\mathbf{q}}(\mathbf{r})}$ where $|c_n(\mathbf{q})|^2$ follows a Gaussian distribution centered at $\mathbf{Q}$.
In our numerical calculations, the coefficient $c_n(\mathbf{q})$ is chosen as
\begin{equation}
c_n(\mathbf{q}) = \mathcal{N}\prod_\mu \exp{\left( -\frac{(q^\mu - Q^\mu)^2}{4\sigma_{Q^\mu}^2} \right)} \,,
\end{equation}
where $\mathcal{N}$ is a normalization constant to make $\int \mathrm{d}^{D}\mathbf{q}~|c_n(\mathbf{q})|^2 = 1$. 
This choice corresponds to a Gaussian wavepacket in momentum space centered at $Q^{\mu}$ with $\sigma_{Q^\mu}$ along each momentum direction $\mu$. However, due to the gauge freedom of the Bloch states, the wavepacket is not uniquely defined~\cite{Marzari-RevModPhys.84.1419}. 
In particular, substituting $\ket{\psi^{(n)}_{\mathbf{q}}}$ with $e^{i \varphi_n(\mathbf{q})}\ket{\psi^{(n)}_{\mathbf{q}}}$, where $\varphi_n(\mathbf{q})$ is a smooth and periodic function in momentum space, results in wavepackets with distinct shapes and spatial spreads. 
As shown by the EOM in the main text, the dynamics of a wavepacket under a nonlinear potential depends not only on its mean position but also on its shape, characterized by quantum correlations.
Therefore, the wavepacket’s spatial structure plays a crucial role in determining its evolution.
Since constructing the wavepacket under an arbitrary gauge choice can result in different dynamical behavior, it is essential to adopt a consistent and physically meaningful gauge.
To this end, we construct a maximally localized wavepacket in real space, which serves as a well-defined benchmark for subsequent calculations.
In our calculations, we first compute the Bloch Hamiltonian $H(\mathbf{q})$, obtain its eigenstates $\ket{u_{n,\mathbf{q}}}$, and then construct the wavepacket state accordingly. 
However, this procedure does not inherently guarantee that the resulting wavepacket is well-localized or symmetric in real space. 
To address this, we adopt the approach developed by Marzari and Vanderbilt for constructing maximally localized Wannier functions~\cite{Marzari-PhysRevB.56.12847}. 
The corresponding localization optimization functional reads
\begin{equation}
\Omega = \sum_{n} [\bra{\psi} \hat{r}^2 \ket{\psi} - \bra{\psi} \hat{r} \ket{\psi}^2] \,,
\end{equation}
where the optimization is performed over the phase degree of freedom $\varphi_n(\mathbf{q})$. 
For our relatively small system sizes ($L \lesssim 300$), a standard gradient descent algorithm is sufficient to minimize $\Omega$ and obtain a well-localized wavepacket.

This method is particularly effective in one-dimensional systems. 
However, in two dimensions, the optimization becomes significantly more demanding due to the increased parameter space and the non-trivial landscape of the functional.


\end{document}